\documentclass[showpacs,twocolumn,showkeys,amsmath,amssymb,prx]{revtex4-2}

\usepackage{bm}
\usepackage{bbm}
\usepackage{dsfont}
\usepackage{color}
\usepackage{graphicx}
\usepackage{cancel}  
\newcommand{\be}{\begin{equation}}
\newcommand{\ee}{\end{equation}}
\newcommand{\rd}{{\mathrm d}}
\newcommand{\re}{{\mathrm e}}
\newcommand{\ri}{{\mathrm i}}

\begin{document}

\title[Floquet dynamics of ultracold atoms in optical lattices with a parametrically modulated trapping potential]
      {Floquet dynamics of ultracold atoms in optical lattices with a parametrically modulated trapping potential}  

\author{Usman Ali$^{1}$, Martin Holthaus$^{2}$, and Torsten Meier$^{1}$}

\affiliation{$^1$Department of Physics, Paderborn University,  
	D-33098 Paderborn, Germany}		
\affiliation{$^2$Institut f\"ur Physik, Carl von Ossietzky Universit\"at, 
	D-26111 Oldenburg, Germany}	
                  
\date{\today}

\begin{abstract}
	
{Experiments with ultracold atoms in optical lattices usually involve a weak parabolic trapping potential which merely serves to confine the atoms, but otherwise remains negligible. In contrast, we suggest a different class of experiments in which the presence of a stronger trap is an essential part of the set-up. Because the trap-modified on-site energies exhibit a slowly varying level spacing, similar to that of an anharmonic oscillator, an additional time-periodic trap modulation with judiciously chosen parameters creates nonlinear resonances which enable efficient Floquet engineering. We employ a Mathieu approximation for constructing the near-resonant Floquet states in an accurate manner and demonstrate the emergence of effective ground states from the resonant trap eigenstates. Moreover, we show that the population of the Floquet states is strongly affected by the phase of a sudden turn-on of the trap modulation, which leads to significantly modified and rich dynamics. As a guideline for further studies, we argue that the deliberate population of only the resonance-induced effective ground states will allow one to realize Floquet condensates which follow classical periodic orbits, thus providing challenging future perspectives for the investigation of the quantum-classical correspondence. }

\end{abstract} 


\maketitle 


\section{Introduction}
\label{S_1}









\par{Quantum wave packets under time-independent Hamiltonians typically exhibit spatial spreading over time due to the presence of dispersion \cite{1}. However, intriguing exceptions exist where wave packets remain localized and follow dynamics similar to those of classical particles. Schr\"odinger was the pioneer in discovering such a quantum system, namely a particle confined within a stationary harmonic potential, where localized wave packets can propagate along classical trajectories without spreading or experiencing dispersion \cite{2}. Another classic example is that of a wave packet in a periodic potential that undergoes coherent Bloch oscillations acted upon by an external constant force \cite{3,4}. It is also known that for these examples, the non-dispersive oscillations turn into breathing dynamics under which the wave packet expands and shrinks periodically \cite{5,6,7,8,9}. Likewise, the coexistence of spreading and localized dynamics has captivated the interest of physicists, leading to extensive research aimed at understanding and harnessing these distinct behaviors \cite{10,11,12,13,14,15,16,17,18,19}. These phenomena hold great potential for applications in metrology, quantum sensing, and imaging, as well as quantum information processing and computing.
\par{Time-periodic driving is a commonly used tool to stabilize the wave packet motion or to modify the irreversible dispersive spread. This is through nonlinear resonances under which non-dispersive wave packets emerge as the eigenstates of the time-dependent system (Floquet states) \cite{20,21,22,23} or as wave packets that exhibit almost perfect recurrence after an integer number of drive periods \cite{24,25,26,27}. These features of the resonant driving have led to predictions of the time crystallinity in homogeneous single-particle lattices \cite{23} as well as in symmetry-breaking many-body systems \cite{28}. At strong driving, delocalized wave packets also emerge, which can coexist with non-dispersive states. In close analogy to the partly chaotic phase space of a classical particle, which presents regular and chaotic motions, the Floquet states can be labeled as regular-resonant and chaotic under the semi-classical eigenfunctions hypothesis \cite{29,30,31,32,33,34}.}
\par{In this paper, we examine the dynamics of wave packets within a periodic potential under the influence of a parametrically modulated parabolic potential. By choosing different initial quantum wave packets we trace the occupation probabilities of Floquet states in the presence of a symmetry-breaking drive. We illustrate the change in occupation probabilities as the time-translation symmetry is broken by a sudden activation of the drive with different phases, or if different initial states are projected onto the Floquet spectrum. We examine the near-resonant Floquet states employing a Mathieu approximation and semiclassical theory, leveraging insights from the classical phase space of the system. As a key result of our analysis, we show in detail that different occupations of Floquet states lead to significantly different dynamics. To demonstrate this, we resort to a one-dimensional (1D) model of a single particle in the one-band tight-binding lattice driven by the parametrically modulated parabolic potential \cite{37,38}. In the stationary limit the model system is no longer described by Bloch bands but instead by localized energy eigenstates with a slowly increasing level spacing, thus mimicking the spectrum of a weakly anharmonic oscillator. It is this anharmonicity which is the precondition for the present study, giving rise to nonlinear resonances when the parametric trap modulation is activated.
Thus, here, the trapping potential does not just provide the wavepacket confinement but is a decisive and constructive element of our considered set-up. The model bears experimental relevance and can be realized with trapped ultracold atoms in optical lattices \cite{41}.

This paper is organized as follows: In Section II, we introduce the model and discuss its classical counterpart. In Section III, we devise the methods for the construction and inspection of different natures of Floquet states. These are shown to possess different occupations upon varying the initial conditions in Section IV. Section V is dedicated to concluding discussions.

\section{The model and its classical counterpart}
\label{S_2}

A quantum particle of mass~$m$ moving in a one-dimensional sinusoidal lattice 
with depth~$V_0$ and lattice constant~$a$ augmented by a parabolic trap 
potential with trapping frequency~$\Omega$ is described by the Hamiltonian   
\be
	H_0 = \frac{p^2}{2m} + V_0 \sin^2\left(\frac{\pi}{a}x\right) 
	+ \frac{1}{2}m\Omega^2x^2 \; . 
\label{eq:HTR}
\ee
If the parabolic trap is modulated periodically in time with frequency~$\omega$
and relative modulation strength~$\alpha$, the total Hamiltonian is given by
\be
	H(t) = H_0 + H_{\rm int}(t) 
\ee
with
\be
	H_{\rm int}(t) = \frac{\alpha}{2}m\Omega^2x^2 \, f(t) \, 
	\sin(\omega t + \phi) \; , 
\label{eq:HIN}
\ee
where the dimensionless function $f(t)$ describes the way the modulation is
turned on within a time interval from $t_i = 0$ to $t_f$, such that
\be
	f(t) = \left\{ \begin{array}{ll}
		0 \; ,	& t < 0	\\
		1 \; ,	& t > t_f \; . 
	\end{array} \right. 
\label{eq:TOF}
\ee
For instance, the extreme case of a sudden turn-on is described by a Heaviside 
function
\be
	f(t) = \Theta(t) \; .
\label{eq:HSF}
\ee
Also note that the specification of the instant $t_i = 0$ as the moment of 
turn-on is what provides physical meaning to the phase~$\phi$ of the modulation.

The essence of the present study already is contained in a single-band, 
nearest-neighbor approximation to the above Hamiltonian $H(t)$. This reduction
leads to 
\be
	\widehat{H}_0 = -J \sum_n \Big( |n+1 \rangle \langle n| 
        + |n \rangle \langle n+1| \Big) 
	+ K_0 \sum_n n^2 |n \rangle \langle n| 
\label{eq:HTB}
\ee
as the remnant of the time-independent operator~(\ref{eq:HTR}), while the
modulation~(\ref{eq:HIN}) furnishes the expression
\be
	\widehat{H}_{\rm int}(t) = \alpha K_0 \, f(t) \, \sin(\omega t + \phi) 
	\sum_n n^2 |n \rangle \langle n| \; .
\label{eq:HDR}
\ee
Here the states $|n\rangle$ denote the Wannier states belonging to the lowest
energy band of the lattice in the absence of the trap, that is, for 
$\Omega = 0$, with $n$ labeling the respective lattice site such that $n=0$
marks the bottom of the parabolic trap. Moreover, $J$ is the hopping matrix
element connecting adjacent sites and $K_0 = m\Omega^2 a^2/2$ is the 
effective trap strength which carries the dimension of an energy.

To make contact with experimentally realistic data and to convey some feeling
for the orders of magnitude of the parameters involved, we consider the 
archetypal case of ultracold $^{87}$Rb atoms in an optical lattice created 
by laser radiation with wavelength $\lambda = 852$~nm~\cite{GreinerEtAl01,
GreinerEtAl02}, implying $a = \lambda/2 = 426$~nm. The customary energy scale 
is then set by the single-photon recoil energy $E_R = \hbar^2\pi^2/(2ma^2)$, 
which is $1.31 \cdot 10^{-11}$~eV under the above conditions. Assuming a 
lattice with a depth of $V_0 = 10~E_R$, the nearest-neighbor approximation 
presupposed in the tight-binding Hamiltonian~(\ref{eq:HTB}) is satisfied to 
the extent that the ratio of the neglected next-to-nearest neighbor hopping 
matrix element to~$J$ is $0.0118$, that is, on the one percent level of 
accuracy~\cite{BoersEtAl07,EckardtEtAl09}. The width of the lowest band 
then amounts to $W = 0.0767~E_R$, so that $J = W/4 = 0.0192~E_R$. Moreover, 
the separation between the lowest two bands figures as $\Delta = 4.57~E_R$. 
In order to uphold the single-band approximation, the trap strength $K_0$ 
should be chosen such that the on-site energy shift $K_0 n^2$ equals the 
band gap~$\Delta$ only for a large number~$n$ of sites, {\em i.e.\/}, for
$n_0 = \sqrt{\Delta/K_0} \gg 1$, but with the trap still exerting a sizeable
influence of the dynamics. Taking $K_0 = 0.32 \times 10^{-3}~E_R$, say, one finds
$n_0 \approx 120$ which is considered to be safe; by virtue of
$\hbar\Omega/E_R = (2/\pi) \sqrt{K_0/E_R}$, this corresponds in our
example to the trapping frequency $\Omega \approx 2\pi \times 36$~Hz.

The driving frequency $\omega$ now should conform to the requirement
$W < \hbar\omega \ll \Delta$, meaning that the drive should be 
non-adiabatically fast in comparison with the band dynamics, but still
induce only negligible interband transitions. Selecting, {\em e.g.\/}, 
$\omega = 2\pi \times 71$~Hz one has $\hbar\omega = 0.0224~E_R$, falling 
well into this window. 

With these caveats, the three dimensionless figures of merit for the model 
defined by $\widehat{H}(t) = \widehat{H}_0 + \widehat{H}_{\rm int}(t)$ are 
$J/(\hbar\omega)$ and $K_0/(\hbar\omega)$ together with the variable relative  
driving strength $\alpha$. As inferred from the scenario envisioned above,
realistic values of $J/(\hbar\omega)$ are on the order of unity or somewhat 
less, whereas $K_0/(\hbar\omega)$ should be about 2 to 3 orders of magnitude 
smaller. The viable range of $\alpha$ again is determined by the requirement 
that interband transitions be avoided; note that $\alpha = 1$, meaning that
the modulation amplitude equals the trap strength itself, already signals strong driving. 

In order to deduce the classical counterpart of the quantum dynamics generated 
by $\widehat{H}(t)$ we consider the position operator 
\be
	\widehat{x} = a \sum_n | n \rangle n \langle n | \; .
\ee
If the label~$n$ were a continuous variable, shifts of position would be
generated by the operator
\be
	\widehat{k} = \frac{1}{\ri} \frac{\rd}{\rd(na)} \; .
\ee
In the case of discrete $n$, as encountered here, the derivative has to be
understood as a finite difference, implying that $\widehat{k}$ is supposed
to act on functions which vary only slowly from site to site. Thus, one has   
\be
	\exp(\ri\widehat{k}a) | n \rangle = | n+1 \rangle \; .
\ee
Introducing the time-dependent trap strength
\be
	K(t) = K_0\big(1 + \alpha f(t)\sin(\omega t + \phi) \big) \; ,
\ee
the Hamiltonian then takes the form
\be
	\widehat{H}(t) = -2J\cos(\widehat{k}a) + K(t) (\widehat{x}/a)^2 \; ,
\ee
where we have made use of the completeness relation
\be
	\sum_n | n \rangle \langle n | = 1 \; .
\ee
For switching from quantum to classical mechanics we now replace the above 
operators $\widehat{x}$ and $\widehat{k}$ by continuous variables $x$ and 
$k$. Since the product $kx$ is dimensionless, whereas the product of two 
canonically conjugate variables should carry the dimension of an action, 
we formally employ the de Broglie relation   
\be
	p = \hbar k \; ,
\ee
leading us to the Hamiltonian function
\be
	H_{\rm cl}(t) = -2J\cos(pa/\hbar) + K(t) (x/a)^2 \; .
\label{eq:PEN}
\ee
Evidently this expression describes a pendulum, with the roles of position and 
momentum interchanged, and endowed with a mass which changes periodically in 
time as $1/K(t)$, providing the equations of motion
\begin{eqnarray}
	\dot{x} & = & \phantom{-}\frac{\partial H_{\rm cl}}{\partial p} 
	\; = \; \frac{2Ja}{\hbar}\sin(pa/\hbar) 
\nonumber \\
	\dot{p} & = & -\frac{\partial H_{\rm cl}}{\partial x} 
	\; = \; -\frac{2}{a^2} K(t) x \; .
\label{eq:HSY}
\end{eqnarray}
According to the Ehrenfest principle~\cite{Ehrenfest27} these classical 
equations~(\ref{eq:HSY}) transform back to quantum mechanics via the 
replacement of $x$ and $p$ by expectation values $\langle \widehat{x} \rangle$ 
and $\hbar\langle \widehat{k} \rangle$. Here we again suppose that these 
expectation values are taken with wave functions which vary slowly and 
smoothly from site to site, which is the very precondition for a semiclassical 
treatment. Thus, one finds  	  
\be
	\frac{\rd}{\rd t} \langle \widehat{x} \rangle = 
	\frac{2Ja}{\hbar}\sin(\langle \widehat{k} \rangle a) 
\label{eq:GVL}
\ee
and
\be
	\hbar \frac{\rd}{\rd t} \langle \widehat{k} \rangle = 
	- \frac{2}{a^2} K(t) \langle \widehat{x} \rangle \; , 
\label{eq:ACT}
\ee
which indeed are meaningful equations: Eq.~(\ref{eq:GVL}) gives the group 
velocity of a wave packet in the cosine energy band provided by the 
tight-binding Hamiltonian~(\ref{eq:HTB}) for $K_0 = 0$, while 
Eq.~(\ref{eq:ACT}) constitutes a tentative generalization of Bloch's 
acceleration theorem~\cite{AliEtAl23}.

\begin{figure}[t]
\centering
\includegraphics[scale=0.265]{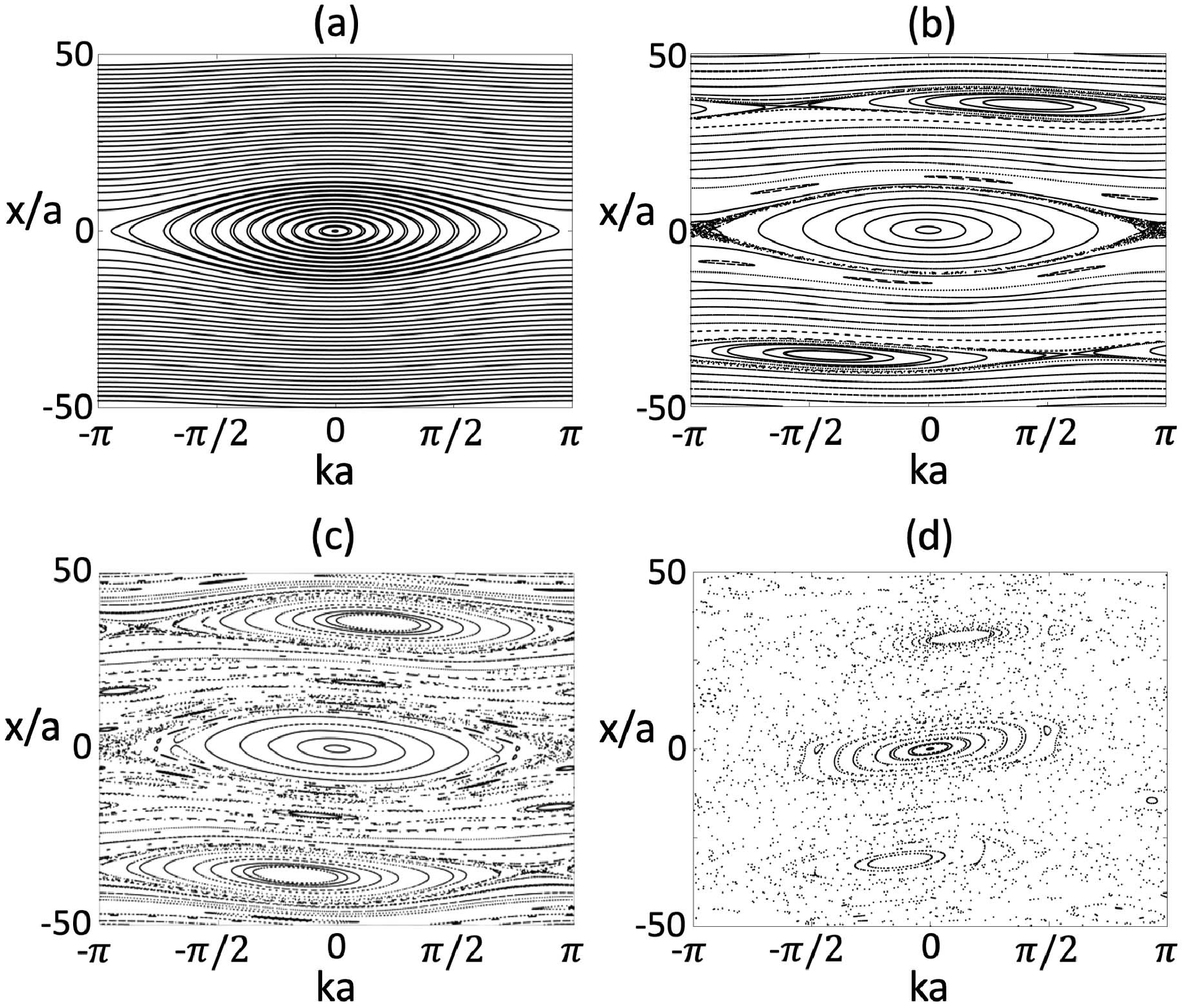}
\caption{Poincar\'e sections pertaining to the classical pendulum 
system~(\ref{eq:HSY}) with $f(t) = 1$, meaning strictly periodic driving, for 
$J/(\hbar\omega) = 1.071$ and $K_0/(\hbar\omega) = 0.0143$. Relative driving 
strengths are $\alpha = 0.01$, $0.25$, $1.00$, and $5.00$ ((a) -- (d)). All 
sections are taken for $t = -\phi/\omega \bmod 2\pi/\omega$. In comparison 
with a standard mechanical pendulum the roles of position and momentum here
are interchanged, in accordance with the underlying Hamiltonian 
function~(\ref{eq:PEN}). Of key importance is the appearance of pendulum-like 
primary resonance zones at $x/a \approx \pm 35$. These are caused by $1:1$ 
resonances between the frequency of oscillations of the undriven pendulum, and 
the frequency of the drive, giving rise to stable oscillations around the new
equilibrium positions.}
\label{F_1}
\end{figure}

As is common practice in the investigation of periodically time-dependent 
Hamiltonian systems, the dynamics of the pendulum system~(\ref{eq:HSY}) are 
visualized by means of stroboscopic Poincar\'e mappings. To this end the 
trajectories which develop from a suitably selected set of initial values 
$(k_i,x_i)$ in the phase space plane are computed numerically for $f(t) = 1$, 
and sampled once per driving period $T = 2\pi/\omega$. In this manner one 
obtains the sections plotted in Fig.~\ref{F_1}, which display characteristic 
features of regular and chaotic classical motion~\cite{LiLi92}. Here we have 
fixed the parameters $J/(\hbar\omega) = 1.071$ and $K_0/(\hbar\omega) = 0.0143$,
but have varied the strength $\alpha$ of the drive. For $\alpha = 0.01$, for 
which $K(t) \approx K_0$, one still finds the familiar phase space portrait of 
an undriven pendulum. Importantly, pronounced pendulum-like primary $1:1$ 
resonance zones have emerged for $\alpha = 0.25$ around $x/a = \pm 35$, 
indicating initial values for which the oscillation frequency of the original 
undriven pendulum closely matches the driving frequency, resulting in stable 
oscillations around the new equilibrium positions. With increasing driving 
amplitude these resonance zones grow larger, while small higher-order resonance 
zones emerge, and chaotic motion becomes more pronounced in the vicinity of the 
separatrices; these mixed regular-chaotic dynamics are depicted exemplarily for 
$\alpha = 1.0$. Increasing the amplitude still further to even $\alpha = 5.0$, 
the $1:1$ resonance zones appear as islands of mainly regular motion embedded 
in a chaotic sea~\cite{LiLi92}.
 
In the following sections we will explore the ramifications of these classical 
near-resonant dynamics for the corresponding quantum system governed by the 
Hamiltonian $\widehat{H}(t)$, that is, for ultracold atoms in optical lattices 
with a periodically modulated trap potential.

\section{Construction and inspection of near-resonant Floquet states}
\label{S_3}
In order to obtain a proper basis for the later discussion of the long-time dynamics initiated by a sudden turn-on of the trap modulation with different phases, we temporarily disregard the actual switch-on process (4) and set $f(t) = 1$. Then the Hamiltonian is strictly periodic in time, $\widehat{H}(t) = \widehat{H}(t+T)$, 
so that it possesses a complete set of Floquet states. These are solutions to 
the time-dependent Schr\"odinger equation of the particular 
form~\cite{Shirley65,Sambe73,Salzman74,BaroneEtAl77,FainshteinEtAl78,Tsuji2024}
\be
	| \psi_j(t) \rangle = | u_j(t) \rangle \,
	\re^{-\ri \varepsilon_j t/\hbar} \; ,
\label{eq:FST}
\ee
combining $T$-periodic Floquet functions
\be
 	| u_j(t) \rangle = | u_j(t+T) \rangle	
\label{eq:TPF}
\ee
with exponentials which specify their time evolution in terms of 
quasienergies~$\varepsilon_j$. The importance of this Floquet basis rests in 
the fact that it allows one to expand every other solution $| \psi(t) \rangle$
to the Schr\"odinger equation with time-independent expansion coefficients 
$c_j$ according to       
\be
	| \psi(t) \rangle = \sum_j c_j \, | \psi_j(t) \rangle \; ,
\label{eq:EXP}
\ee
providing occupation probabilities $|c_j|^2$ which remain constant in time 
despite the periodic driving, and thus characterize the full quantum dynamics 
for all times. While these occupation probabilities will be shown to be 
dependent on the phase of the drive in Sec.~IV, here 
we outline the approximate construction of the Floquet states themselves under 
the near-resonant conditions considered in the previous Sec.~\ref{S_2}. To this 
end we adapt a quantum mechanical resonance analysis which is the analog of a 
classical pendulum approximation~\cite{Chirikov79,BermanZaslavsky77}. Similar 
investigations have been performed before in the contexts of, among others, 
periodically driven quantum wells~\cite{Holthaus95}, subharmonic response 
to time-periodic excitation~\cite{HolthausFlatte94}, generalized 
$\pi$-pulses~\cite{HolthausJust94}, quantum revivals in periodically driven systems~\cite{Fsaif2001,Fsaif2011}, and many-particle tunneling in periodically 
driven Bosonic Josephson junctions~\cite{GertjerenkenHolthaus14}.

The starting point of this analysis is the numerical solution to the energy
eigenvalue problem posed by the tight-binding Hamiltonian~(\ref{eq:HTB}) in 
the presence of the parabolic trap, but without its periodic modulation,   
\be
	\widehat{H}_0 | \varphi_\ell \rangle = 	
	E_\ell | \varphi_\ell \rangle \; .
\label{eq:EEP}
\ee
The eigenstates $\{ |\varphi_\ell\rangle \}$ fall into two groups: Those with 
low energies are localized in the trap center, whereas the high-energy states 
are localized in a Wannier-Stark-like manner around those positions $\pm n_0 a$
where their eigenvalues approximately equal the on-site energy~$K_0 n_0^2$ \cite{37,39,40}.   
By symmetry, the latter appear in pairs with almost degenerate eigenvalues,
indicating long-range tunneling. Here we restrict ourselves to the 
non-degenerate states associated with only one wing of the parabolic trap,
and consider their coupling to the other wing in a second step.

In general, the expansion of a Floquet state~(\ref{eq:FST}) with 
quasienergy~$\varepsilon$ with respect to this restricted energy basis possesses 
the form 
\be
	| \psi(t) \rangle = \re^{-\ri \varepsilon t/\hbar}
	\sum_\ell b_\ell(t) \, | \varphi_\ell \rangle  
\ee
with periodically time-dependent coefficients
\be
	b_\ell(t) = b_\ell(t+T) \; ,
\label{eq:GEC}
\ee
where we have omitted the Floquet state label~$j$ previously employed in 
Eq.~(\ref{eq:FST}) for ease of notation. Since the eigenvalues $E_\ell$ 
increase regularly with their index~$\ell$, the level spacing can be
regarded as a discrete derivative, $E_{\ell + 1} - E_\ell \equiv E'_\ell$.
The quantum analog of the classical $1:1$ resonance condition between 
unperturbed oscillation frequency and driving frequency then is expressed by 
the relation 
\be
	E'_r = \hbar\omega \; ,
\label{eq:RES}
\ee
assumed to be satisfied by a certain ``resonant'' state labeled by the 
index~$r$. Moreover, the expansion coefficients~(\ref{eq:GEC}) accompanying
near-resonant states should be dominated by a single Fourier component,
motivating the Ansatz  
\be
	b_\ell(t) = \widetilde{a}_\ell \, \re^{-\ri(\ell - r)\omega t} \; .
\ee
Thus, the coefficient $b_r$ becomes time-independent, the coefficients 
$b_{r\pm 1}$ are proportional to $\exp(\mp\ri\omega t)$, and so on. For later 
convenience we also set
\be
	\widetilde{a}_\ell = a_\ell \, \re^{-\ri(\ell-r)(\phi - \pi/2)} \; ,
\label{eq:LCO}
\ee
with $a_\ell$ still unknown. In summary, the near-resonant Floquet states 
should adhere approximately to expressions of the form
\be
	| \psi(t) \rangle = \re^{-\ri \varepsilon t/\hbar} \sum_\ell a_\ell \, 
	\re^{-\ri(\ell-r)(\omega t + \phi - \pi/2)} | \varphi_\ell \rangle \; ,
\label{eq:ANS}
\ee
with the implicit understanding that the sum over $\ell$ extends over a limited 
set of states around the resonant one, also requiring $r \gg 1$. Inserting this 
Ansatz~(\ref{eq:ANS}) into the time-dependent Schr\"odinger equation, one 
derives a system of coupled equations which determines both the desired 
coefficients~$a_\ell$ and the quasienergy~$\varepsilon$, namely,
\begin{widetext}
\be
	\big( \varepsilon + (m - r)\hbar\omega \big) a_m 
	= E_m a_m + K_0\alpha \sin(\omega t + \phi) \sum_{n,\ell} 
	\langle \varphi_m | n^2 | n \rangle \langle n | \varphi_\ell \rangle 
	\, a_\ell \, \re^{\ri(m-\ell)(\omega t + \phi - \pi/2)} \; .
\label{eq:COE}
\ee
\end{widetext}
This suggests a combination of three further approximations. First, in order 
to facilitate the required time-independence of the coefficients only the 
secular terms with $\ell = m \pm 1$ are kept, as corresponding to the familiar 
rotating-wave approximation. Second, the matrix elements which couple 
neighboring coefficients have to be approximated by a common constant
\be
	K_0\alpha \sum_n \langle \varphi_m | n^2 | n \rangle 
	\langle n | \varphi_{m\pm 1} \rangle \equiv V \; .
\label{eq:COM}
\ee 
Third, the energy eigenvalues adjacent to the resonant one are expanded
quadratically, giving
\be
	E_m = E_r + (m-r)\hbar\omega + \frac{1}{2} (m-r)^2 E_r'' \; ,
\ee
having employed the resonance condition~(\ref{eq:RES}) and relying on the fact 
that the level spacing increases only slowly with~$m$. Taken together, these 
steps lead us from Eqs.~(\ref{eq:COE}) to the strongly simplified system
\be
	\varepsilon a_m = E_r a_m + \frac{1}{2}(m - r)^2 E_r'' a_m
	+ \frac{V}{2} \big(a_{m+1} + a_{m-1} \big) \; .
\label{eq:SSS}
\ee
Now regarding the $a_m$ as Fourier coefficients of a function $\chi(z)$
according to the prescription
\be
	a_m = \frac{1}{\pi} \int_0^\pi \! \rd z \; \chi(z) \,
	\re^{\ri(m - r)2z} \; ,
\label{eq:FOC}
\ee
one has  
\begin{eqnarray}
	& & (m-r)^2 a_m \phantom{\int}
\nonumber \\ & = &
	\frac{1}{\pi} \int_0^\pi \! \rd z \; \chi(z)
	\left( -\frac{1}{4} \frac{\rd^2}{\rd z^2} \re^{\ri(m - r)2z} \right)
\nonumber \\ & = &
	\frac{1}{\pi} \int_0^\pi \! \rd z  
        \left( -\frac{1}{4} \frac{\rd^2}{\rd z^2} \chi(z) \right)
	\re^{\ri(m - r)2z}  
\end{eqnarray}
with the proviso that $\chi(z)$ be a $\pi$-periodic function, 
$\chi(z) = \chi(z+\pi)$, so that the partial integrations under\-lying the 
second equality do not produce boundary terms. Thus, the system~(\ref{eq:SSS})
is transformed into the differential equation
\be
	(\varepsilon - E_r) \chi(z) = -\frac{1}{8}E''_r \chi''(z)
	+ V\cos(2z) \chi(z) \; ,
\label{eq:MAT}
\ee
which is recognized as the stationary Schr\"odinger equation for a fictitious 
particle with an effective mass proportional to $1/E''_r$ in a cosine 
potential, {\em i.e.\/}, for a pendulum. 

This new pendulum is not directly 
related to the undriven Hamiltonian~(\ref{eq:HTB}), the  classical analog of 
which gives rise to the phase space diagram depicted in Fig.~\ref{F_1}(a), but 
rather to the quantum counterparts of the pendula which manifest themselves in 
Figs.~\ref{F_1}(b)-(d) through the resonance zones at $x/a \approx \pm 35$. 
While this approximate construction is valid only ``locally'' in the vicinity
of the resonant eigenstate and thus yields merely the near-resonant Floquet 
states, it has the merit of converting the Floquet problem into a much simpler 
energy eigenvalue problem for an undriven quantum pendulum.  
  
\begin{figure}[t]
\centering
\includegraphics[scale=0.4]{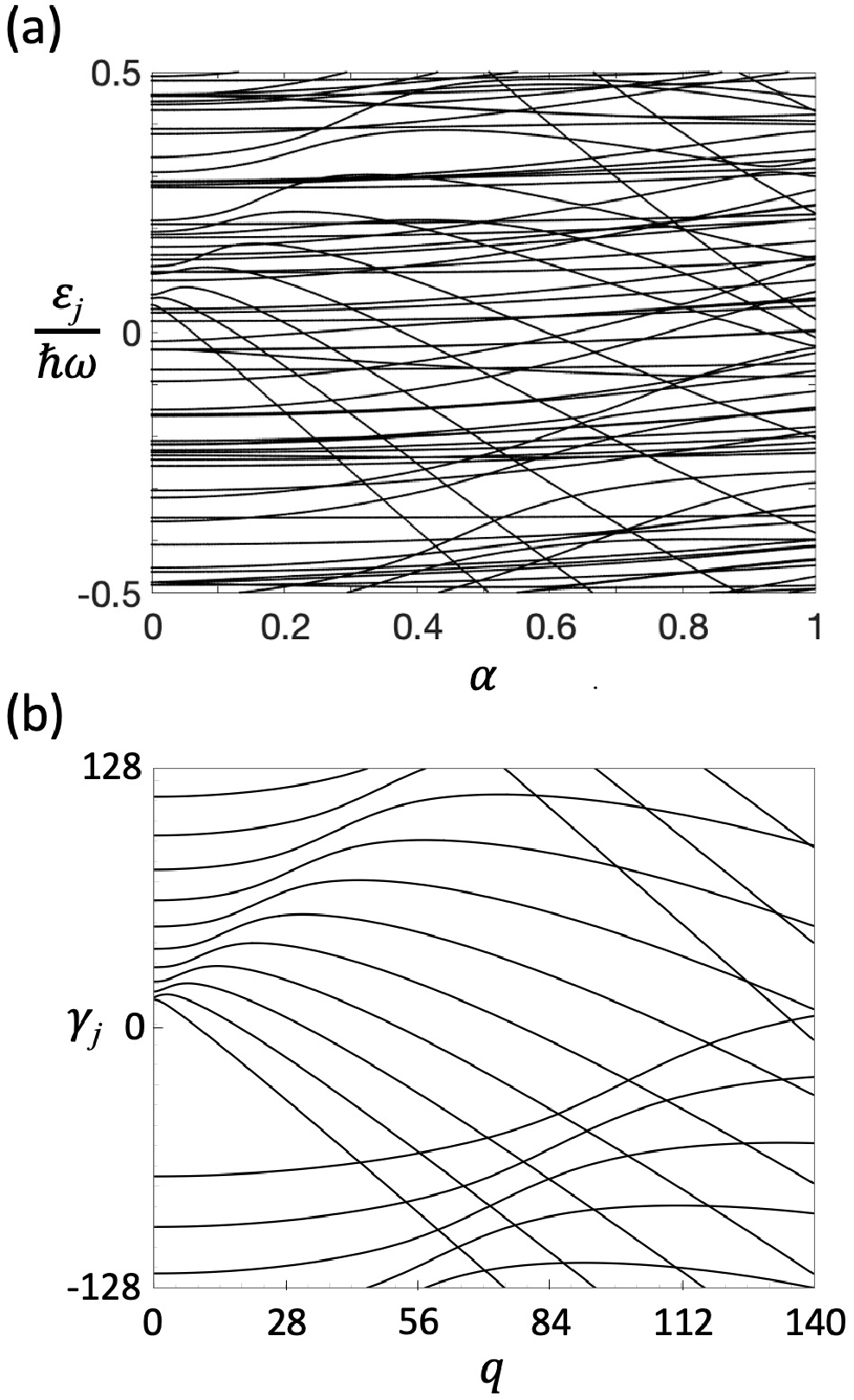}
\caption{(a): One Brillouin zone of exact, numerically computed quasienergies 
for the Hamiltonian $\widehat{H}_0 + \widehat{H}_{\rm int}(t)$, as given by
Eqs.~(\ref{eq:HTB}) and~(\ref{eq:HDR}) with $f(t) = 1$. All quasienergies 
are taken modulo~$\hbar\omega$. Parameters are $J/(\hbar\omega) = 1.071$ 
and $K_0/(\hbar\omega) = 0.0143$, as in Fig.~\ref{F_1}. Quasienergies of 
near-resonant states are two-fold (almost) degenerate, corresponding to 
the two wings of the parabolic trapping potential. Quasienergies of the 
resonance-induced Floquet ground-state doublet $j=0$ are designated by the 
line with the strongest descent. 
(b) Approximate quasienergies according to Eq.~(\ref{eq:NRQT}) for the same 
parameters.}
\label{F_2}
\end{figure}  

The eigenvalue equation~(\ref{eq:MAT}) has the form of a Mathieu equation, 
which is canonically written as~\cite{AS72} 
\be
	y'' + \big[a - 2q \cos(2z) \big] y = 0 \; ;
\label{eq:NOF}
\ee
here we have the parameters 
\begin{eqnarray}
	a & = & \frac{8(\varepsilon - E_r)}{E''_r}
\nonumber \\
	q & = & \frac{4V}{E''_r} \; .
\label{eq:MAP}
\end{eqnarray}
Now we can resort to well-known particular properties of the Mathieu equation. 
Given any value of the depth~$q$ of the effective cosine well, which according 
to Eq.~(\ref{eq:COM}) is proportional to the driving amplitude~$K_0\alpha$, 
the required $\pi$-periodic functions $\chi(z)$ exist only for certain 
so-called characteristic values of the other parameter~$a$; this is the 
quantization condition which yields the quasienergies~$\varepsilon$. These 
characteristic values are denoted by $a_{2r}$, associated with even 
$\pi$-periodic Mathieu functions, and $b_{2r}$, associated with odd such 
functions~\cite{AS72}. Using this nomenclature, and defining
\be
	\gamma_j = \left\{ \begin{array}{ll} 
	a_{j} \; ,	& j = 0,2,4,\ldots \\
	b_{j+1} \; ,	& j = 1,3,5,\ldots \; ,
	\end{array} \right.
\label{eq:NRQT}
\ee
the first Eq.~(\ref{eq:MAP}) gives the quasienergies of the near-resonant 
Floquet states~(\ref{eq:ANS}) in the form
\be
	\varepsilon_j = E_r + \frac{1}{8}E''_r \gamma_j
\ee
with quantum number $j = 0,1,2,\ldots\,$; their expansion coefficients~$a_\ell$
finally are provided by the Fourier coefficients~(\ref{eq:FOC}) of the 
corresponding Mathieu functions $\chi_j(z)$. In hindsight, the use of the 
variable $2z$ in Eq.~(\ref{eq:FOC}) serves to arrive directly at the normal 
form~(\ref{eq:NOF}) of the Mathieu equation; likewise, the particular 
factorization~(\ref{eq:LCO}) merely serves to synchronize the Floquet states 
$|\psi_j(t)\rangle$ with the phase of the trap modulation.   

In view of the series of approximations involved in the derivation of the 
near-resonant Floquet states and their quasienergies one might doubt the 
accuracy of the resulting expressions. Such doubts are dispelled by 
Fig.~\ref{F_2}, in which we compare the exact, numerically computed 
quasienergies to the prediction~(\ref{eq:NRQT}) for the same para\-meters 
as previously employed in Fig.~\ref{F_1}. Since the above analysis applies
separately to each wing of the trapping parabola, the actual Floquet states
appear in pairs, approximated by even and odd linear combinations of states 
obtained by the Mathieu construction, so that the exact quasienergies are 
two-fold (almost) degenerate, with a minuscule tunneling splitting not visible 
in Fig.~\ref{F_2}(a). Other than that, the analytical approximation depicted in 
Fig.~\ref{F_2}(b) works even unexpectedly well indeed, providing additional 
support for the following deliberations.

A path-directing finding implied by the expression~(\ref{eq:NRQT}) is the 
emergence of a Floquet-state quantum number~$j = 0,1,2,\ldots$ which is 
different from the quantum number~$\ell$ which labels the solutions to the 
energy eigenvalue problem~(\ref{eq:EEP}) pertaining to the undriven trap. 
Once again, the physical meaning of this new quantum number is clarified 
by semiclassical considerations. Namely, Floquet states which correspond 
to regular classical motion inside the resonant islands showing up in 
Fig.~\ref{F_1} are associated with invariant tubes in the extended phase space
$\{ (k,x,t) \}$; these tubes are obtained by following the Hamiltonian flow 
of the closed curves which surround the elliptic fixed point in the centers 
of the islands. Those classical tubes which ``carry'' a quantum Floquet state 
are singled out by the Bohr-Sommerfeld-like condition~\cite{Gutzwiller90, BreuerHolthaus91}            
\be
\oint_{\gamma_j} \! k \, \rd x = 2\pi\left( j + \frac{1}{2} \right) \; ,
\label{eq:BHQ}
\ee
with integer $j = 0,1,2,\ldots$ and paths $\gamma_j$ which wind once around
the respective tube. Thus, the state with $j = 0$ is associated with the 
innermost quantized tube and therefore may be regarded as a {\em bona fide\/} 
ground state; the state with $j = 1$ is associated with the next tube matching 
the hierarchy~(\ref{eq:BHQ}) and serves as a first excited state, and so on. 
In effect, the integer~$j$ entering the quantization rule~(\ref{eq:BHQ}) 
equals the quantum number~$j$ introduced in Eq.~(\ref{eq:NRQT}). Moreover, 
Eq.~(\ref{eq:BHQ}) testifies that a Floquet state claims an area of $2\pi$ in 
the $\{(k,x)\}$ phase space plane at each instant of time. Thus, the resonant 
islands showing up in Fig.~\ref{F_1}(c), taken together, should host about 
$10$~pairs of almost degenerate Floquet states, coupled by dynamical tunneling 
through the zone of mixed regular-chaotic motion in between, as is well 
confirmed by Fig.~\ref{F_2}.

The appearance of resonance-induced effective Floquet ground states $j = 0$ 
can also be demonstrated in a purely quantum mechanical manner. Here a 
physics-based conceptual complication comes into play, arising from the 
innocuous-looking identity        
\be
	| u(t) \rangle \, \re^{-\ri \varepsilon t/\hbar}  =
        | u(t) \re^{\ri m \omega t} \rangle \, 
	\re^{-\ri (\varepsilon + m\hbar\omega) t/\hbar} \; . 	
\ee
This means that the factorization of a given Floquet state~(\ref{eq:FST}) into 
a $T$-periodic Floquet function~(\ref{eq:TPF}) and an accompanying exponential 
is not unique; the Floquet function~$| u(t) \rangle$ represents the very same 
Floquet state as does $| u(t) \re^{\ri m \omega t} \rangle$ for any 
$m = 0, \pm 1, \pm 2, \ldots$ Likewise, a quasienergy should not be regarded 
as a single quantity~$\varepsilon$, but rather as a ladder of equally spaced 
representatives $\varepsilon + m\hbar\omega$. Thus, the quasienergy spectrum is 
unbounded from both above and below, with one representative of the quasienergy 
of each Floquet state falling into each Brillouin zone of width~$\hbar\omega$. 
As a consequence, Floquet states admit no meaningful ordering with respect to 
the magnitude of their quasienergies. While this well-known fact appears to be 
without relevance in the regime of perturbative weak driving, where the 
Floquet states can be unambiguously connected to the unperturbed energy 
eigenstates~$|\varphi_\ell\rangle$ from which they descend, this is no longer
the case for strong driving. Here we resort to the following procedure: 
We compute all Floquet states $|u(t_0)\rangle\exp(-\ri\varepsilon t_0/\hbar)$ 
at an instant $t_0 = -\phi/\omega$ at which the drive~(\ref{eq:HDR}) vanishes, 
and compute the instantaneous expectation values 
$\langle u(t_0) | \widehat{H_0} | u(t_0) \rangle$. The Floquet states then 
are ordered according to the magnitude of these expectation values, such that 

\be
\langle u_\ell(t_0) | \widehat{H_0} | u_\ell(t_0) \rangle \leq
\langle u_{\ell+1}(t_0) | \widehat{H_0} | u_{\ell+1}(t_0) \rangle
\label{eq:LKO}
\ee
for $\ell = 0,1,2,\ldots$ While this {\em ad hoc\/} procedure is not 
perfect, it offers the distinct merit of being free of arbitrariness.

\begin{figure}[t]
\centering
\includegraphics[scale=0.4]{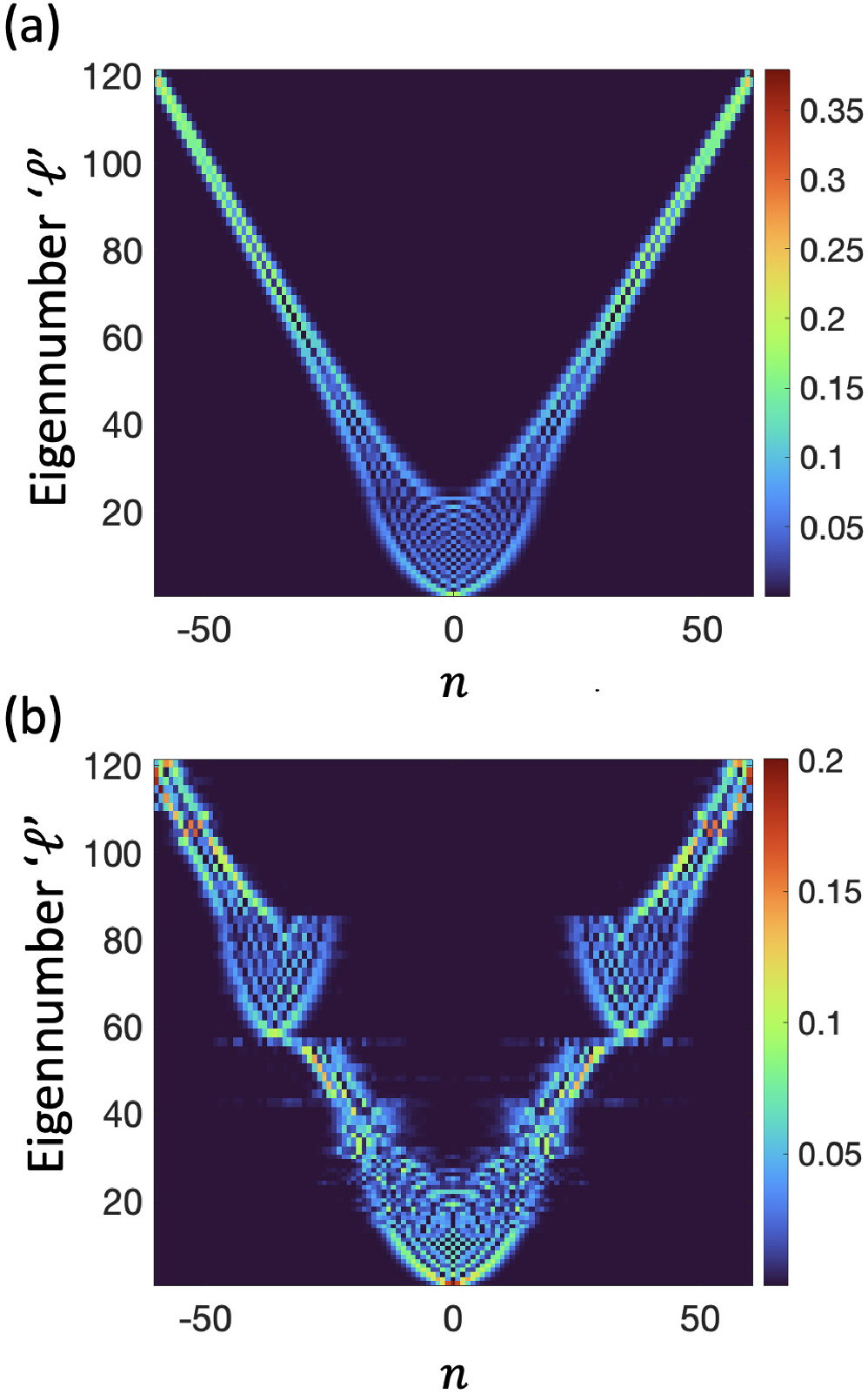}
\caption{Color-coded occupation probabilities of a Wannier state 
$|n \rangle$ located at the $n$-th lattice site as provided by (a) the unperturbed energy eigenstates through $p(n,\ell) = | \langle n |\varphi_\ell \rangle |^2$, and (b) the squared-overlap $p(n,\ell) = | \langle n |u_{\ell}(t_0) |^2$ with the Floquet states $|u_{\ell}(t)\rangle\exp(-\ri\varepsilon_{\ell} t/\hbar)$ at an instant~$t_0$ at which the drive~(\ref{eq:HDR}) vanishes, with the Floquet-state labels~$\ell$ being determined by the instantaneous energy ordering according to Eq.~(\ref{eq:LKO}). Parameters are the same as in Fig.~\ref{F_1}(c). Observe the appearance of a resonance-induced ground state doublet for $\ell = r = 58$ at $n = \pm 35$.}
\label{F_3}
\end{figure} 

In Fig.~\ref{F_3} we present a comparison between the eigenstates of the unperturbed system and a particular visualization of the Floquet states, which is based on the above auxiliary ordering. Therein we plot the color-coded occupation probabilities of the Wannier states $| n \rangle$ effectuated by the squared overlap with the unperturbed energy eigenstates $|\varphi_{\ell}\rangle$ as shown in Fig.~\ref{F_3}(a) and with the Floquet states $|u_{\ell}(t_0)\rangle\exp(-\ri\varepsilon_{\ell} t_0/\hbar)$ as depicted in Fig.~\ref{F_3}(b), with $t_0$ as above, for the same parameters as used before in Fig.~\ref{F_1}(c). Fig.~\ref{F_3}(a) exhibits the harmonic oscillator-like localized states in the trap center and the Wannier-Stark-like strongly localized states at positions $\pm n_0 a$ for eigennumbers above a critical energy index, which is given by $\ell_c=2||2J/K_0||$, where $||y||$ represents the rounding off to the closest integer to $y$. The former are semiclassically associated with the elliptic closed curves, shown in Fig.~\ref{F_1}(a), indicating oscillating pendulum motion, the latter with the curves above the separatrix. Similarly, in Fig.~\ref{F_3}(b), for low $\ell$ ranging from $\ell = 0$ to about $\ell = 15$ one observes harmonic oscillator-like Floquet states which belong to the central island depicted in Fig.~\ref{F_1}(c). Here the likeness of the Floquet states to the unperturbed trap eigenstates is still quite pronounced, so that our ordering works well. In the intermediate range $15 < \ell < 58$ the joint images of the Floquet states appear somewhat blurred, corresponding to the intricate classical motion showing up between the regular islands in Fig.~\ref{F_1}(c). The appearance of two branches reflects the localization of the states at the two turning points of classical motion, corresponding to almost degenerate pairs of Floquet states localized around both points. At $\ell = r = 58$, marking the resonant energy eigenstate, the quantum counterparts of the resonant islands make themselves felt, resulting in further harmonic oscillator-like states placed in both wings of the trapping parabola at $n = \pm 35$, matching precisely the centers of the islands. These are the states covered by the Mathieu approximation, with a resonance-induced ground state doublet $j = 0$ emerging from the resonant energy eigenstate $\ell = 58$. Evidently, the Mathieu hierarchy of near-resonant Floquet states manifesting itself here comprises about $20$~states, once again in fair accordance with the combined areas of the $1:1$ resonant islands in Fig.~\ref{F_1}(c). 

\begin{figure*}[t]
\centering
\includegraphics[scale=0.35]{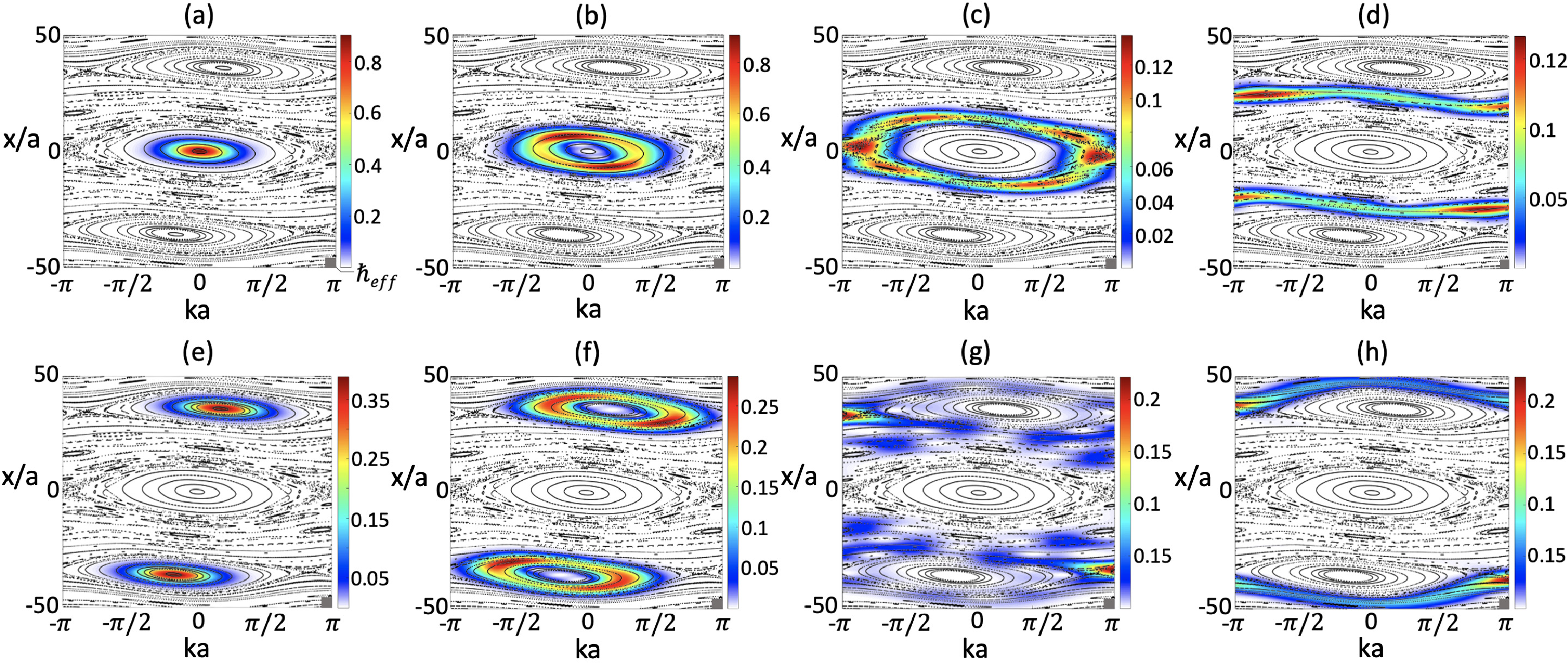}
\caption{Color-coded Floquet-state Husimi distributions~(\ref{eq:HDF}) 
superimposed on the Poincar\'e section taken from Fig.~\ref{F_1}(c), with 
parameters as stated there. With respect to the instantaneous energy ordering 
also employed in Fig.~\ref{F_3}, the Floquet states portrayed in the present 
figure carry the labels $\ell = 0, 10, 15, 45, 58, 65, 85$, and $86$ 
((a) - (h)). Recall that $\ell = r = 58$ (panel~(e)) corresponds to $j = 0$, 
indicating a member of the resonance-induced ground state doublet.}
\label{F_4}
\end{figure*}

To conclude our discussion of the quantum-classical correspondence for cold
atoms in an optical lattice with a periodically modulated parabolic trap
potential, we display in Fig.~\ref{F_4} Husimi distributions of some especially
characteristic Floquet states with parameters as above, superimposed on the 
Poincar\'e section already fleshed out in Fig.~\ref{F_1}(c). These color-coded 
distributions quantify the squared overlaps
\be
	Q_\ell(z) = \big| \langle z | u_\ell(t_0) \rangle \big|^2 
\label{eq:HDF}
\ee
of a Floquet state $|u_\ell(t_0)\rangle\exp(-\ri\varepsilon_\ell t_0/\hbar)$ 
which is sliced at $t_0 = -\phi/\omega$ in order to match the stroboscopic 
recording of the classical trajectories, with coherent states $| z \rangle$ 
which are positioned at points $(k,x)$ in the phase space plane by setting 
$z = x/a + \ri ka$~\cite{Schleich01}. Panels~(a) and~(b) refer to $\ell = 0$ 
and $\ell = 10$, respectively, portraying the Floquet states which stem 
from the ground state and from the $10$-th exited state of the undriven 
Hamiltonian~(\ref{eq:HTB}); both states still fit into the central regular 
island. Panel~(c) depicts the state labeled $\ell = 15$ by virtue of the 
instantaneous energy ordering~(\ref{eq:LKO}); this state appears to be 
associated with both its broken separatrix and a higher-order island chain. 
Panel~(d) then shows a Floquet state which corresponds to a superposition of 
left- and rightward rotations of the pendulum~(\ref{eq:PEN}); these translate 
into vibrations of the actual particle in the left and right wing of the trap. 
Now focusing on the $1:1$ resonant islands, panel~(e) shows a member of the 
resonance-induced ground-state doublet $j = 0$, while panel~(f) depicts an 
excited state of the Mathieu hierarchy. As expected, these states concentrate 
on the elliptical-shaped closed curves which are selected by the quantization 
condition~(\ref{eq:BHQ}); their simultaneous occupation of both resonant 
islands indicates long-range quantum tunneling. In contrast, panel~(g) displays 
a Floquet state which is mainly associated with the chaotic dynamics outside 
the islands, whereas panel~(h) visualizes another rotational state. 

Thus, one may roughly divide the Floquet states into ``regular'', 
``resonant-regular'', and ``chaotic'' ones. Consequently, in a laboratory
experiment one may expect wave packet dynamics which depend substantially on 
the category of states that are populated in its initial stage.

\section{Floquet-state occupation probabilities for sudden turn on}}
\label{S_4}

Having inspected the Floquet basis states we now return to the central 
Eq.~(\ref{eq:EXP}) and consider the expansion coefficients~$c_\ell$ which
provide the Floquet-state occupation probabilities~$|c_\ell|^2$. As already
described these coefficients are determined by the way the trap modulation is
switched on, as specified by the envelope function~(\ref{eq:TOF}). Here,
we restrict ourselves to a sudden turn-on modeled by the Heaviside 
function~(\ref{eq:HSF}) and will establish standards for the Floquet engineering options that later could be complemented with more general, deliberately designed smooth turn-on processes. Although such an instantaneous onset of the trap modulation might not seem realistic from an experimental point of view, it may serve to illustrate some salient features. Furthermore, situations which are essentially described well by an instantaneous turn have been realized in experiment, see e.g.,~\cite{Cao2020} where a significantly modified transport dynamics which is controlled by the phase of the drive has been demonstrated.

Given an initial wave packet $|\psi(0)\rangle$ at $t = 0$, taken to be the 
moment of a sudden turn-on, the expansion coefficients are provided by the 
projection of this initial state onto the Floquet states, such that
\be
	c_\ell = \langle u_\ell(0) | \psi(0) \rangle \; .
\ee 
If we synchronize the Floquet states with the argument $\omega t + \phi$ of 
the drive~(\ref{eq:HDR}), writing $|\widetilde{u}_\ell(\omega t + \phi)\rangle$ 
instead of $|u_\ell(t)\rangle$, it becomes evident that choosing different 
phases~$\phi$ at $t=0$ is equivalent to sampling the Floquet states at 
different instants of their evolution, so that the expansion coefficients 
depend on~$\phi$.

\begin{figure}[t]
\centering
\includegraphics[scale=0.22]{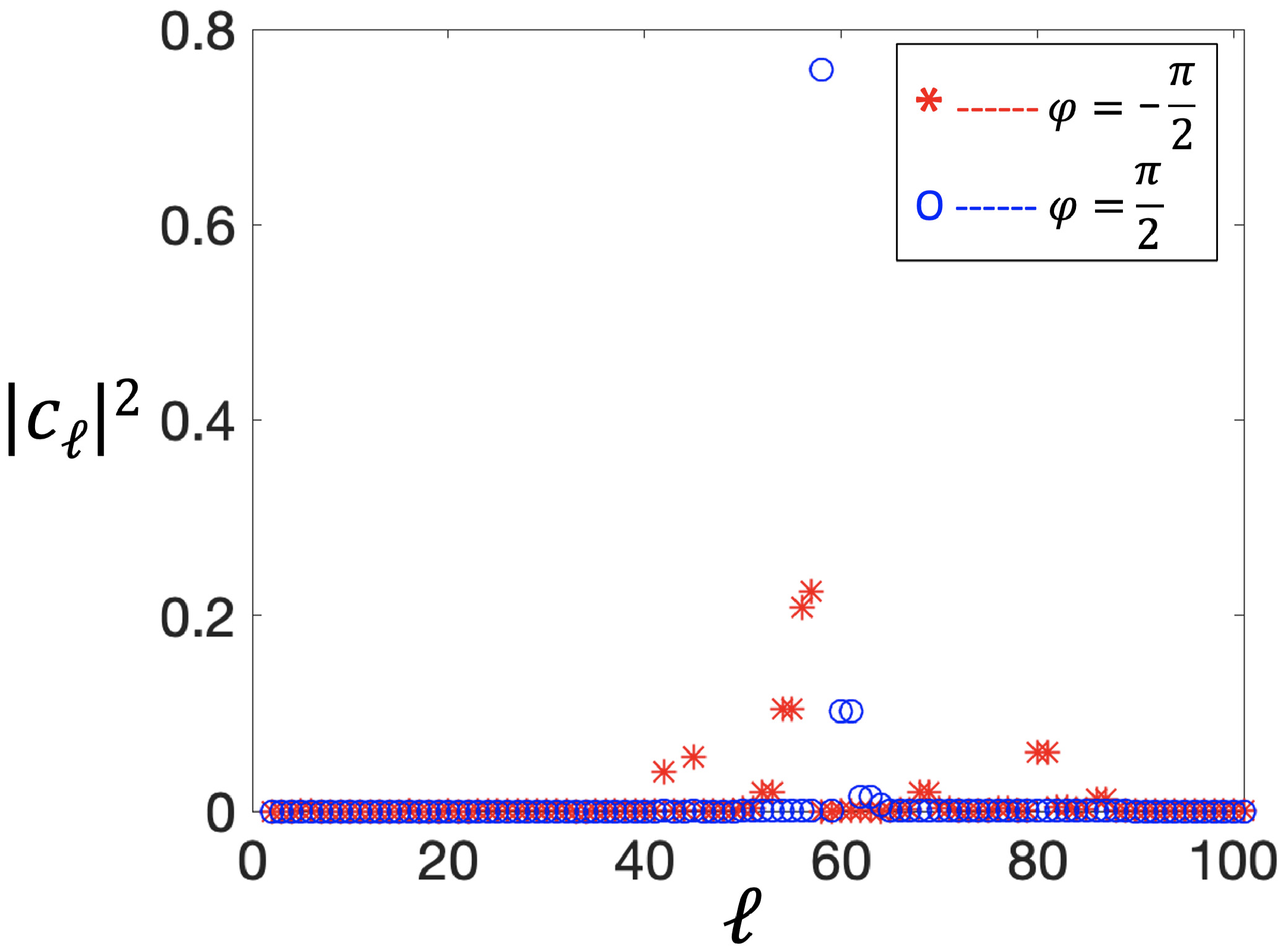}
\caption{Floquet-state occupation probabilities $|c_\ell|^2$ obtained after a sudden turn-on of the trap modulation for initial Gaussian wave packets~(\ref{eq:GWP}) with width~$\sigma = 2.23$, centered at $n_0 = 35$, with $\phi = \pm\pi/2$. As before, Floquet-state labels~$\ell$ are assigned according to the instantaneous energy ordering brought about by Eq.~(\ref{eq:LKO}). Once more, parameters are fixed as in Fig.~\ref{F_1}(c).}
\label{F_5}
\end{figure}

For demonstration purposes we select Gaussian packets
\be
	|\psi(0)\rangle = \sum_n \frac{1}{\sqrt{\sigma\pi}}
	\exp\!\left(-\frac{(n - n_0)^2}{2\sigma^2}\right) | n \rangle 
\label{eq:GWP}
\ee
as initial states, centered with zero momentum at the lattice site~$n_0$. We 
consider fairly narrow packets with a width~$\sigma$ of $2.23$ lattice sites
only, once again take the same parameters as in Fig.~\ref{F_1}(c), and plot 
in Fig.~\ref{F_5}(a) the occupation probabilities $|c_\ell|^2$  obtained for 
$n_0 = 35$, as corresponding to the center of the $1:1$ resonant islands, 
for both $\phi = +\pi/2$ and $\phi = -\pi/2$. In the former case only a few 
Floquet states are populated, with a significantly high population appearing
at $\ell = r = 58$. In addition one finds pairs of states carrying almost
identical probabilities; these are doublets of symmetry-related Floquet states 
associated with the regular resonant islands. In contrast, for $\phi = -\pi/2$ 
the total probability is shared among several more Floquet states; these are 
identified as separatrix states or adjacent ones. 

\begin{figure}[t]
\centering
\includegraphics[scale=0.495]{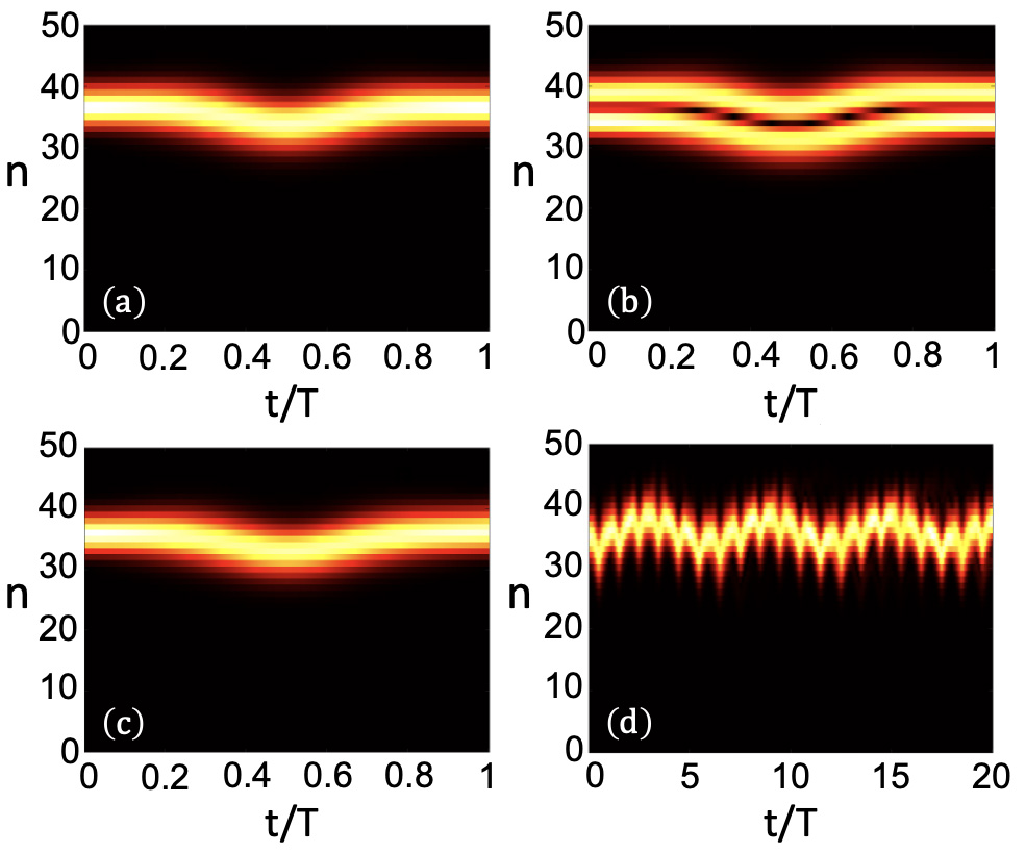}
\caption{Absolute values of the one-period evolution of the maximally occupied Floquet states corresponding to $\ell = r = 58$ in (a) and $\ell = 60$ in (b) for the case with $\phi = -\pi/2$. (c) One period evolution generated by the superposition of all the occupied Floquet states and (d) the corresponding long-time dynamics.}
\label{F_6}
\end{figure}

Next, to illustrate the emerging dynamics, we present the periodic evolution of selected highly occupied Floquet states in Figs.~\ref{F_6} and \ref{F_7}. Additionally, the quasiperiodic solutions resulting as the superposition of all the Floquet states are also displayed. Specifically, in Fig.~\ref{F_6}(a) and (b), we showcase the absolute value of the one-period evolution for the Floquet states corresponding to $\ell = r = 58$ and $\ell = 60$. These states are identified as carrying the highest population in Fig.~\ref{F_5} for the case of $\phi = -\pi/2$. Moreover, these states are seen to manifest the Mathieu hierarchy where Fig.~\ref{F_6}(a) and (b) are a member of the ground and first excited state doublet $j=0$ and $1$, respectively. Further, the one-period evolution exhibits a slow oscillation of these states which is the reminiscent of the Bloch-like dynamics on the arms of the parabolic lattice~\cite{AliEtAl23, Ponokolo2006, Brandkolo2007}. Figure~\ref{F_6}(c) captures the one-period evolution generated by the initial Gaussian wave packet (\ref{eq:GWP}), which represents the superposition of all the occupied Floquet states. The dynamics reveals no noticeable difference in comparison to the dynamics of the effective Floquet ground state due to a strong occupation of this state. However, Fig.~\ref{F_6}(d) demonstrates the corresponding long-time dynamics, offering insights into unnoticeably small differences over a period that lead to a net wave packet transport across the lattice. Given the occupation of only a few Floquet states belonging to the $1:1$ resonant island, the transport dynamics maintain coherence and a sub-harmonic motion is generated with a period equivalent to the period of regular trajectories inside the resonant island.

\begin{figure}[t]
\centering
\includegraphics[scale=0.47]{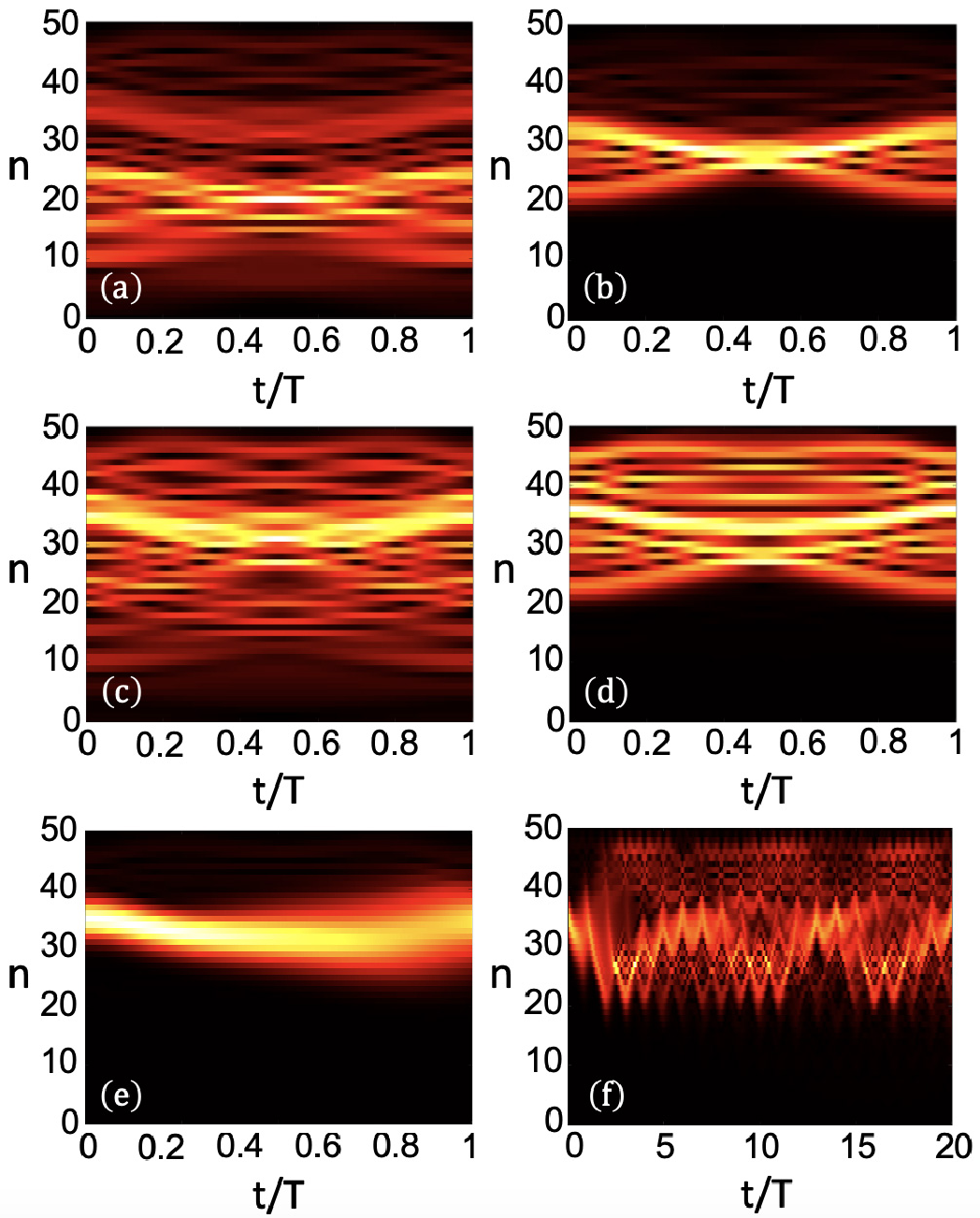}
\caption{Absolute values of the one-period evolution of the maximally occupied Floquet states corresponding to $\ell = 45$ in (a), $\ell = 54$ in (b), $\ell = 57$ in (c) and $\ell = 81$ in (d), for the case with $\phi = \pi/2$. (e) One period evolution generated by the superposition of all the occupied Floquet states and (f) the corresponding long-time dynamics.}
\label{F_7}
\end{figure}

In Fig.~\ref{F_7}(a)-(d) the absolute values of the one-period evolution of the Floquet states corresponding to $\ell = 45$, $54$, $57$, {and} $81$ for the case of $\phi = \pi/2$ are displayed. These are the separatrix states which belong to the broken separatrix around the regular resonant island at the positive axis. The state in Fig.~\ref{F_7}(a) shows quite involved mixed dynamics. However, the one in Fig.~\ref{F_7}(b) exhibits breathing motion along the separatrix. Similar to Fig.~\ref{F_7}(a), the state in Fig.~\ref{F_7}(c) possesses several modes and the dynamics are very complex. The phase space dynamics of these two states (not shown) resemble Fig.~\ref{F_4}(g) and thus they are perceived as chaotic states. Fig.~\ref{F_7} (d) displays a highly excited state of the resonance-induced effective pendulum, which fall around the separatrix. Keeping in view the high population of separatrix states a nonuniform spreading motion is anticipated when we superimpose all the states. Accordingly, an easily perceivable spreading motion is revealed for the propagated Gaussian wave packet, shown in Fig.~\ref{F_7}(e). The quasiperiodic evolution seen here leads to further spreading and recombinations which gives rise to quite complicated dynamics, as shown in Fig.~\ref{F_7}(f). These dynamics represent a combination of Bloch-breathing and -oscillatory modes, showcasing the persistent Bloch-like oscillatory character in both the above examples. However, the variation in the dynamics is borne out in the diverse nature of states emerging due to external periodic driving. By the virtue of the drive phase the occupation of these states can be controlled.

A simplistic viewpoint on the origin of varying occupation numbers can be made by slicing the Poincar\'e surfaces of sections at times equivalent to the drive phase. This is demonstrated in Fig.~\ref{F_8} where we plot the Poincar\'e phase space at $\omega t = \pm \pi/2$ and the regular resonant islands appear at different positions. The Poincar\'e sections are superimposed on the Husimi distribution of the initial Gaussian state (\ref{eq:GWP}). Fig.~\ref{F_8}(a) displays the intersection of the initial Gaussian wave packet with the regular resonant island for $\phi = -\pi/2$. Thus, the occupation of near-resonant Floquet states for this phase is highlighted. Similarly, for $\phi = \pi/2$, Fig.~\ref{F_8}(b) illustrates the activation of separatrix states as the initial Gaussian wave packet superposes with the separatrix trajectory at the hyperbolic fixed point. 

\begin{figure}[t]
\includegraphics[scale=0.09]{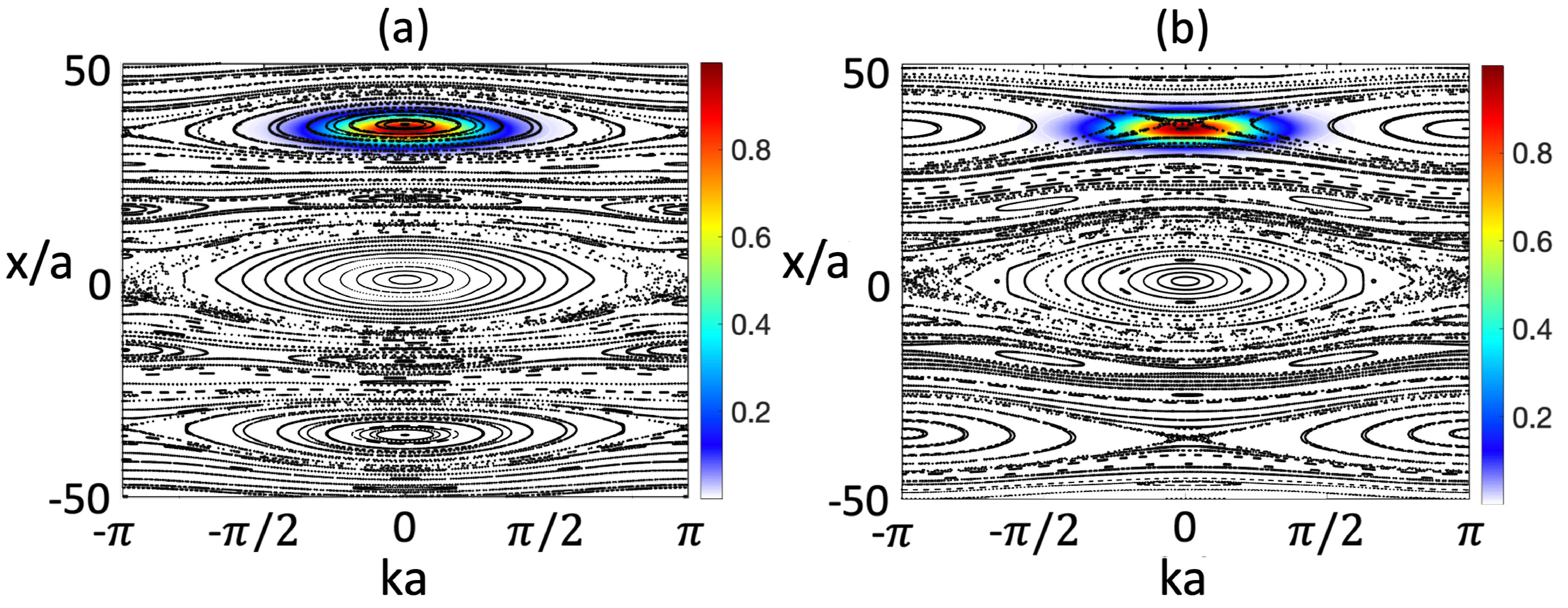}
\caption{Color-coded Husimi distribution~(\ref{eq:HDF}) of the initial state (\ref{eq:GWP}) superimposed on the Poincar\'e sections traced under the opposite parity drive with $\phi = -\pi/2$ (a) and $\phi = \pi/2$ (b). All other parameters remain the same as in Fig.~\ref{F_1}(c).}
\label{F_8}
\end{figure}



\begin{figure}[b]
\centering
\includegraphics[scale=0.2]{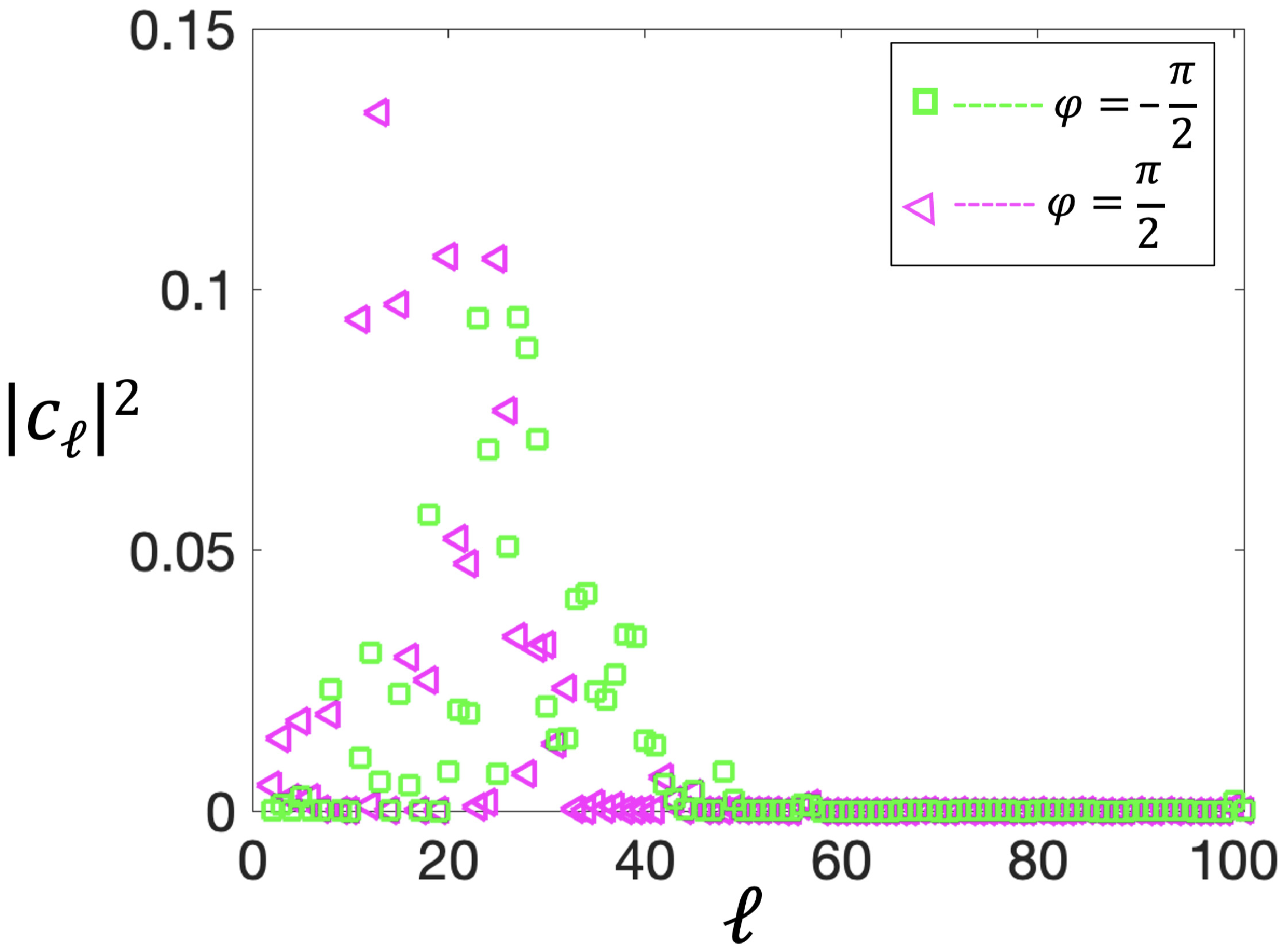}
\caption{Floquet-state occupation probabilities $|c_\ell|^2$ obtained after a sudden turn-on of the trap modulation for initial Gaussian wave packets~(\ref{eq:GWP}) with width~$\sigma = 2.23$, centered at $n_0 = 17$, with $\phi = \pm\pi/2$. As before, Floquet-state labels~$\ell$ are assigned according to the instantaneous energy ordering brought about by Eq.~(\ref{eq:LKO}). Once more, parameters are fixed as in Fig.~\ref{F_1}(c).}
\label{F_9}
\end{figure}

\begin{figure}[t]
\hspace*{-0.3cm} 
\includegraphics[scale=0.5]{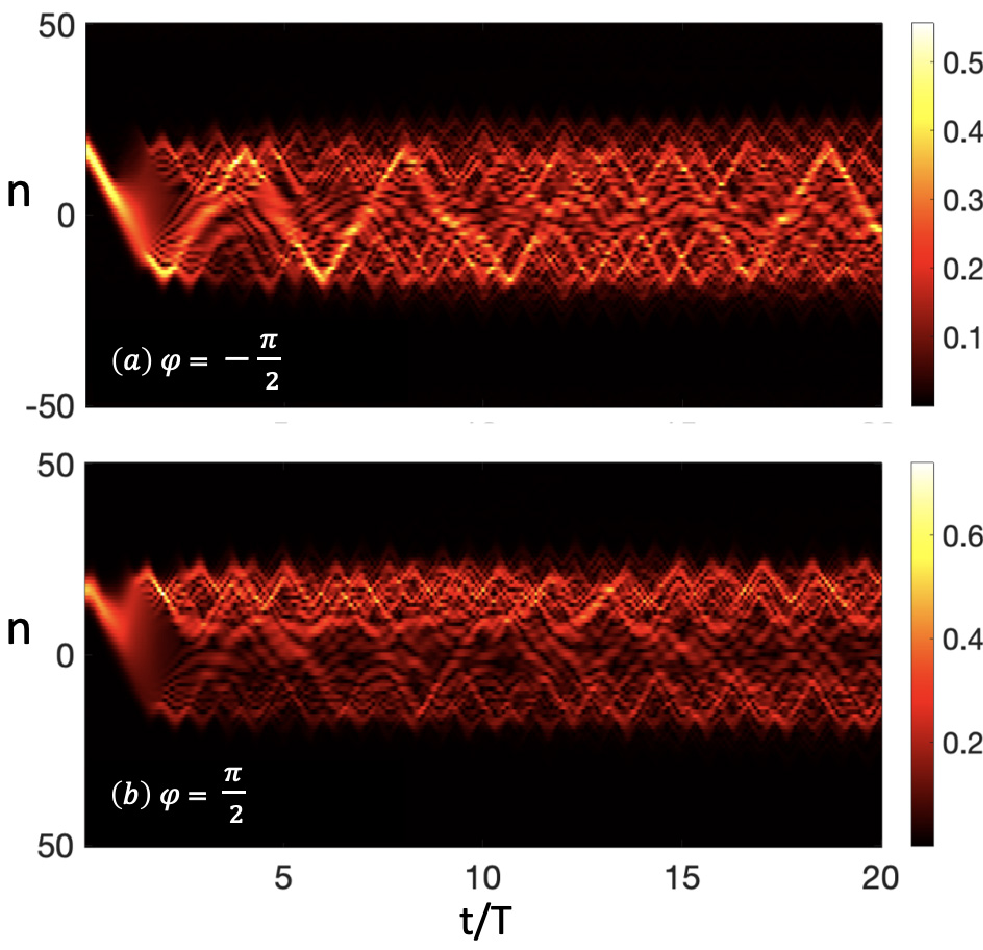}
\caption{Absolute value of the wave packet evolution for the initial Gaussian placed at the lattice location $n_0=17$ with width $\sigma_0=2.23$ under resonant driving for opposite drive phases of $\phi = -\pi/2$ (a) and $\phi = \pi/2$ (b). All other parameters remain the same as in the previous figures.}
\label{F_10}
\end{figure}

As an additional result, Fig.~\ref{F_9} displays occupation probabilities for $n_0 = 17$, falling right at the broken separatrix at $k=0$. Here, the occupation probabilities for both $\phi = \pm \pi/2$ do not follow a discernible pattern. Also, many more states are populated for these examples due to the separatrix breaking at the on-set of chaos. Moreover, the high occupation is on the states in the intermediate regime, i.e., $15<\ell<58$, which belong to the intricate regions of mixed dynamics between the regular islands. The wave packet dynamics corresponding to these occupation probabilities are depicted in Fig.~\ref{F_10}. Here, the initial wave packet soon spreads about symmetrically around the trap center $n = 0$ in a seemingly irregular fashion. Here the chaotic nature of the occupied Floquet states manifests itself, extending through all of the chaotic zone between the islands. Of course, these coherent long-time dynamics are strictly quasiperiodic, in particular for such wave packets which consist of only a limited number of states, but the return times might be far longer than the duration of an actual experiments.






\vspace{2cm}

\section{Conclusion}
\label{S_6}
Ultracold atoms in optical lattices often are employed to simulate phenomena occurring in condensed-matter physics, such as the archetypal superfluid-to-Mott insulator transition \cite{GreinerEtAl02}. In contrast, the present investigation suggests to divert activities into a different direction. Implementing a parabolic trapping potential with non-negligible strength and thus converting the Bloch bands of the lattice into a set of localized energy eigenstates with slowly varying level spacing one effectively realizes an anharmonic oscillator, with an anharmonicity which can be externally controlled by suitably adjusting the trap strength. When that trapping potential is modulated periodically in time, with a frequency which matches the anharmonic level spacing of energy eigenstates well above the original ground state, nonlinear resonances appear in the phase space of the system's classical analog, as exemplified by Fig. 1 of the present study. Because the resonant zones of predominantly regular classical motion cover areas which are sufficiently large to support several quantum mechanical Floquet states, as singled out by the Bohr-Sommerfeld-like quantization condition (40), a distinct hierarchy of near-resonant Floquet states establishes itself, in accordance with the quantum number $j$ showing up in Eqs. (39) and (40). This semiclassical reasoning, forecasting the emergence of resonance-induced effective ground states, is borne out in a purely quantum mechanical manner by the Mathieu approximation, and convincingly confirmed without any approximation by Fig. 3. Here one clearly observes the semiclassically predicted re-organization of states, together with the transmutation of resonant energy eigenstates into effective Floquet ground states.

A major part of our study has been devoted to the question how different populations of resonance-induced regular, separatrix-dominated, or chaotic Floquet states determine the long-time evolution of the driven ultracold atoms. In practice, these populations are determined by the way the periodic trap modulation is turned on. Using the customary example of a sudden turn-on we have demonstrated that the phase of this turn-on may serve as an efficient means to manipulate the occupations of the various types of Floquet states, thereby effectuating significantly different dynamics which should be discernible in currently feasible laboratory experiments. Altogether, our findings suggest that ultracold atoms in optical lattices with a periodically modulated parabolic trapping potential may serve as a promising platform for unvealing novel aspects of the classical-quantum correspondence. 

There is a particularly tantalizing direction of future research which now almost suggests itself. With specifically designed protocols (4) for the turn-on of the periodic drive it will be feasible to populate exclusively the resonance-induced effective Floquet ground states. When working with weakly interacting Bose-Einstein condensates this will amount to the macroscopic population of a Floquet state which follows the classical periodic orbit provided by the Poincar\'e-Birkhoff fixed point theorem, thereby endowing a genuinely quantum-mechanical many-body system with classical dynamics. In the light of our observations, this scenario appears to lie well within the bounds of possibility.

\begin{acknowledgments}
U.A.~gratefully acknowledges support from Deutscher Akademischer 
Austauschdienst (DAAD, German Academic Exchange Service) through a doctoral research grant and we thank the PC$^2$ (Paderborn Center for Parallel Computing) for providing computing time. M.H.~has been supported by the Deutsche Forschungsgemeinschaft (DFG, German Research Foundation) through Project No.~397122187. 
\end{acknowledgments}


\begin{thebibliography}{1}

\bibitem{1}
A. Messiah, Quantum Mechanics, Vol. \textbf{1} (North-Holland Publishing Co., Amsterdam, 1964).
\bibitem{2}
E. Schr\"odinger, Der stetige \"Ubergang von der Mikro- zur Makromechanik, Naturwissenschaften \textbf{14}, 664 (1926).
\bibitem{3}
F. Bloch,  \"Uber die Quantenmechanik der Elektronen in Kristallgittern, Z. Phys. \textbf{52}, 555 (1929).
\bibitem{4}
C. Zener, A theory of the electrical breakdown of solid dielectrics, Proc. R. Soc. London A \textbf{145}, 523 (1934).
\bibitem{5}
V. Grecchi and A. Sacchetti, Acceleration theorem for Bloch oscillators, Phys. Rev. B \textbf{63}, 212303 (2001).
\bibitem{6}
A. M. Bouchard and M. Luban, Bloch oscillations and other dynamical phenomena of electrons in semiconductor superlattices, Phys. Rev. B \textbf{52}, 7 (1995).
\bibitem{7}
T. Hartmann, F. Keck, H. J. Korsch, and S. Mossmann, Dynamics of Bloch oscillations, New J. Phys. \textbf{6}, 2 (2004).
\bibitem{8}
C. C. Guerry and P.L. Knight, Introductory Quantum Optics (Cambridge University Press, Cambridge, 2005), pp. 150-165.
\bibitem{9}
G. Breitenbach, S. Schiller, and J. Mlynek, Measurement of the quantum states of squeezed light, Nature (London) \textbf{387}, 471 (1997).
\bibitem{10}
M. Weiss, T. Kottos, and T. Geisel, Spreading and localization of wavepackets in disordered wires in a magnetic field, Phys. Rev. B \textbf{63}, 081306(R) (2001)
\bibitem{11}
P. E. Grabowski, A. Markmann, I. V. Morozov, I. A. Valuev, C. A. Fichtl, D. F. Richards, V. S. Batista, F. R. Graziani, and M. S. Murillo, Wave packet spreading and localization in electron-nuclear scattering, Phys. Rev. E \textbf{87}, 063104 (2013)
\bibitem{12}
R. Grobe and M. V. Fedorov, Wavepacket spreading and electron localization in strong-field ionization, J. Phys. B: At. Mol. Opt. Phys. \textbf{26}, 1181 (1993)
\bibitem{13}
M. F. Herman and  D. F. Coker, Classical mechanics and the spreading of localized wave packets in condensed phase molecular systems, J. Chem. Phys. \textbf{111}, 1801 (1999)
\bibitem{14}
F. A. Cuevas, S. Curilef, and A. R. Plastino, Spread of highly localized wave-packet in the tight-binding lattice: Entropic and information-theoretical characterization, Ann. Phys. \textbf{326}, 2834 (2011)
\bibitem{15}
M. Yessenov, J. F., Z. Chen, E. G. Johnson, M. P. J. Lavery, M. A. Alonso, and A. F. Abouraddy, Space-time wave packets localized in all dimensions, Nat. Commun. \textbf{13}, 4573 (2022)
\bibitem{16}
R. Stuetzle, M. C. Goebel, T. Hoerner, E. Kierig, I. Mourachko, M. K. Oberthaler, M. A. Efremov, M. V. Fedorov, V. P. Yakovlev, K. A. H. van Leeuwen, and W. P. Schleich, Observation of Nonspreading Wave Packets in an Imaginary Potential, Phys. Rev. Lett. \textbf{95}, 110405 (2005)
\bibitem{17}
B. P. Nguyen, Q. M. Ngo, and K. Kim, Transient super-ballistic spreading of wave packets with large spreading exponents in some hybrid ordered-quasiperiodic lattices, J. Korean Phys. Soc. \textbf{68}, 387 (2016)
\bibitem{18}
M. Amini, Spread of wave packets in disordered hierarchical lattices, Eu. Phys. Lett. \textbf{117}, 30003 (2017)
\bibitem{19}
M. O. Sales, W. S. Dias, A. R. Neto, M. L. Lyra, and F. A. B. F. de Moura, Sub-diffusive spreading and anomalous localization in a 2D Anderson model with off-diagonal nonlinearity,  Solid State Commun. \textbf{270}, 6 (2018)
\bibitem{20}
M. Holthaus, On the classical-quantum correspondence for periodically time dependent systems, Chaos, Solitons \& Fractals \textbf{5}, 1143 (1995).
\bibitem{21}
H. Maeda and T. F. Gallagher, Nondispersing Wave Packets, Phys. Rev. Lett. \textbf{92}, 133004 (2004).
\bibitem{22}
M. Kalinski, L. Hansen, and D. Farrelly, Nondispersive Two-Electron Wave Packets in a Helium Atom, Phys. Rev. Lett. \textbf{95}, 103001 (2005).
\bibitem{23}
Z. Huang, A. Clerk, and I. Martin, Nondispersing Wave Packets in Lattice Floquet Systems, Phys. Rev. Lett. \textbf{126}, 100601 (2021).
\bibitem{24}
A. Buchleitner, D. Delande, and J. Zakrzewski, Non-dispersive wave packets in periodically driven quantum systems, Phys. Rep. \textbf{368}, 409 (2002).
\bibitem{25}
A. Buchleitner and D. Delande, Nondispersive Electronic Wave Packets in Multiphoton Processes, Phys. Rev. Lett. \textbf{75}, 1487 (1995).
\bibitem{26}
L. V. Vela-Arevalo and R. F. Fox, Coherent states of the driven Rydberg atom: Quantum-classical correspondence of periodically driven systems, Phys. Rev. A \textbf{71}, 063403 (2005).
\bibitem{27}
A. Goussev, P. Reck, F. Moser, A. Moro, C. Gorini, and K. Richter, Overcoming dispersive spreading of quantum wave packets via periodic nonlinear kicking, Phys. Rev. A \textbf{98}, 013620 (2018).
\bibitem{28}
K. Sacha, Modeling spontaneous breaking of time-translation symmetry, Phys. Rev. A \textbf{91}, 033617 (2015).
\bibitem{29}
B. Gertjerenken and M. Holthaus, Trojan quasiparticles, New J. Phys. \textbf{16}, 093009 (2014).
\bibitem{30}
M. Martinez, O. Giraud, D. Ullmo, J. Billy, D. Gue\'ry-Odelin, B. Georgeot, and G. Lemari\'e, Chaos-Assisted Long-Range Tunneling for Quantum Simulation, Phys. Rev. Lett. \textbf{126}, 174102 (2021).
\bibitem{31}
M. Arnal, G. Chatelain, M. Martinez, N. Dupont, O. Giraud, D. Ullmo, B. Georgeot, G. Lemarie\', J. Billy, and D. Gue\'ry-Odelin, Chaos-assisted tunneling resonances in a synthetic Floquet superlattice, Sci. Adv. \textbf{6}, eabc4886 (2020).
\bibitem{32}
R. Dubertrand, J. Billy, D. Gu\'ery-Odelin, B. Georgeot, and G. Lemari\'e, Routes towards the experimental observation of the large fluctuations due to chaos-assisted tunneling effects with cold atoms, Phys. Rev. A \textbf{94}, 043621 (2016).
\bibitem{33}
K. W. Mahmud, H. Perry, and W. P. Reinhardt, Quantum phase-space picture of Bose-Einstein condensates in a double well, Phys. Rev. A \textbf{71}, 023615 (2005).
\bibitem{34}
R. Ketzmerick, and W. Wustmann, Statistical mechanics of Floquet systems with regular and chaotic states, Phys. Rev. E \textbf{82}, 021114 (2010).
\bibitem{35}
O. R. Diermann, Mathieu-state reordering in periodic thermodynamics, Z. Naturforsch. \textbf{76}, 12 (2021).
\bibitem{36}
A. Eckardt and E. Anisimovas, High-frequency approximation for periodically driven quantum systems from a Floquet-space perspective, New. J. Phys. \textbf{17}, 093039 (2015).
\bibitem{37}
U. Ali, M. Holthaus, and T. Meier, Chirped Bloch-harmonic oscillations in a parametrically forced optical lattice, Phys. Rev. Research \textbf{5}, 043152 (2023).
\bibitem{38}
N. Mann, M. Reza Bakhtiari, F. Massel, A. Pelster, and M. Thorwart, Driven Bose-Hubbard model with a parametrically modulated harmonic trap, Phys. Rev. A \textbf{95}, 043604 (2017).
\bibitem{39}
A. M. Rey, G. Pupillo, C. W. Clark, and C. J. Williams, Ultracold atoms confined in an optical lattice plus parabolic potential: A closed-form approach, Phys. Rev. A \textbf{72}, 033616 (2005).
\bibitem{40}
D. McKay, M. White, and B. DeMarco, Lattice thermodynamics for ultracold atoms, Phys. Rev. A \textbf{79}, 063605 (2009).
\bibitem{41}
M. M. Neiczer, Efficient creation of a molecular Bose-Einstein condensate of Lithium-6 using a spatially modulated dipole trap, Master Thesis, University of Heidelberg (2018).

\bibitem{GreinerEtAl01}
	{M. Greiner, I. Bloch, O. Mandel, T. W. H\"ansch, and T. Esslinger,
	{Exploring Phase Coherence in a 2D Lattice of Bose-Einstein
	Condensates},
	Phys. Rev. Lett. {\bf 87}, 160405 (2001).}

\bibitem{GreinerEtAl02}
	{M. Greiner, O. Mandel, T. Esslinger, T. W. H\"ansch, and I. Bloch,
	{Quantum Phase Transition from a Superfluid to a Mott Insulator
	in a Gas of Ultracold Atoms},
	Nature {\bf 415}, 39 (2002).}

\bibitem{BoersEtAl07}
	{D. J. Boers, B. Goedeke, D. Hinrichs, and M. Holthaus,
	{Mobility edges in bichromatic optical lattices},
	Phys. Rev. A {\bf 75}, 063404 (2007).}

\bibitem{EckardtEtAl09}
	{A. Eckardt, M.~Holthaus, H. Lignier, A. Zenesini, D. Ciampini,
	O. Morsch, and E. Arimondo,
	{Exploring dynamic localization with a Bose-Einstein condensate},
	Phys Rev. A {\bf 79}, 013611 (2009).}

\bibitem{Ehrenfest27}
	{P. Ehrenfest, 
	{Bemerkung \"uber die angen\"aherte G\"ultigkeit der klassischen 
	Mechanik innerhalb der Quantenmechanik}, 
	Z. Phys. {\bf 45}, 455 (1927).}

\bibitem{AliEtAl23}
	U. Ali, M. Holthaus, and T. Meier,
	{Chirped Bloch-harmonic oscillations in a parametrically forced
	optical lattice},
	Phys. Rev. Research {\bf 5}, 043152 (2023). 

\bibitem{LiLi92}
A. J. Lichtenberg and M. A. Lieberman, Regular and Chaotic Motion (Springer, New York, 1992), Applied Mathematical Sciences (AMS) \textbf{38}.

\bibitem{Shirley65}
{	J. H. Shirley,
	{Solution of the Schr\"odinger Equation with a Hamiltonian 
	Periodic in Time},
	Phys. Rev. {\bf 138}, B 979 (1965). }

\bibitem{Sambe73}
{	H. Sambe,
	{Steady States and Quasienegies of a Quantum-Mechanical System 
	in an Oscillating Field},
	Phys. Rev.~A {\bf 7}, 2203 (1973).}

\bibitem{Salzman74}
{	W. R. Salzman,
	{Quantum Mechanics of Systems Periodic in Time},
	Phys. Rev.~A {\bf 10}, 461 (1974).}

\bibitem{BaroneEtAl77}
{	S. R. Barone, M. A. Narcowich, and F. J. Narcowich,
	{Floquet Theory and Applications},
	Phys. Rev.~A {\bf 15}, 1109 (1977).}

\bibitem{FainshteinEtAl78}
	{A. G. Fainshtein, N. L. Manakov, and L. P. Rapoport,
	{Some General Properties of Quasi-Energetic Spectra of
	Quantum Systems in Classical Monochromatic Fields},
	J.~Phys.~B: Atom. Molec. Phys. {\bf 11}, 2561 (1978).}

\bibitem{Tsuji2024}
N. Tsuji, Encyclopedia of Condensed Matter Physics (Academic press/Elsevier, Amsterdam, Netherlands, 2024), pp. 967-980.

\bibitem{Chirikov79}
	{B. V. Chirikov,
	{A Universal Instability of Many-Dimensional Oscillator systems},
	Phys. Rep. {\bf 52}, 263 (1979). }

\bibitem{BermanZaslavsky77}	
	{G. P. Berman and G. M. Zaslavsky,
	{Theory of Quantum Nonlinear Resonance}, 
	Phys. Lett. A {\bf 61}, 295 (1977).}

\bibitem{Holthaus95}
	{M. Holthaus,
	{On the Classical-Quantum Correspondence for Periodically Time
	Dependent Systems},
	Chaos, Solitons, \& Fractals {\bf 5}
	(Special issue: {Quantum Chaos: Present and Future}),
	1143 (1995).}

\bibitem{HolthausFlatte94}
{	M. Holthaus and M. E. Flatt\'{e},
	{ Subharmonic Generation in Quantum Systems},
	Phys. Lett. A {\bf 187}, 151 (1994).	}	 

\bibitem{HolthausJust94}
	{M. Holthaus and B.~Just,
        { Generalized $\pi$ Pulses},
	Phys. Rev. A {\bf 49}, 1950 (1994).	}

\bibitem{Fsaif2001}
F. Saif and M. Fortunato, Quantum revivals in periodically driven systems close to nonlinear resonances, Phys. Rev. A \textbf{65}, 013401 (2001).

\bibitem{Fsaif2011}
M. Ayub, K. Naseer, and F. Saif, Robust dynamical recurrences based on Floquet spectrum, Eur. Phys. J. D \textbf{64}, 491 (2011).	

\bibitem{GertjerenkenHolthaus14}
	{B. Gertjerenken and M. Holthaus,
	{Trojan Quasiparticles},
	New J. Phys. {\bf 16}, 093009 (2014).}

\bibitem{AS72}
M. A. Abramowitz and I. A. Stegun, Handbook of Mathematical Functions, (Dover Publications, Inc, New York, 1972), chap. \textbf{20}.	
\bibitem{Gutzwiller90}
M. C. Gutzwiller, Chaos in Classical and Quantum Mechanics (Springer, New York, 1990).

\bibitem{BreuerHolthaus91}
	H. P. Breuer and M. Holthaus,
	{A Semiclassical Theory of Quasienergies and Floquet Wave 
	Functions},
	Ann. Phys. (N.Y.) {\bf 211}, 249 (1991).

\bibitem{Schleich01}
	{W. P. Schleich,
	{Quantum Optics in Phase Space},
	(Wiley-VCH, Berlin, 2001).}
	

\bibitem{Cao2020}
A. Cao, R. Sajjad, E. Q. Simmons, C. J. Fujiwara, T. Shimasaki, and D. M. Weld, Transport controlled by Poincar\'e orbit topology in a driven inhomogeneous lattice gas, Phys. Rev. Res. \textbf{2}, 032032(R) (2020).
	

\bibitem{Ponokolo2006}
A. V. Ponomarev and A. R. Kolovsky, Dipole and Bloch oscillations of cold atoms in a parabolic lattice, Laser Physics, \textbf{16}, 367 (2006).

\bibitem{Brandkolo2007}
J. Brand and A. R. Kolovsky, Emergence of superfluid transport in a dynamical system of ultracold atoms, Eur. Phys. J. D, \textbf{41}, 331 (2007).
	












		  	
\end{thebibliography}
\end{document}